\documentclass[aps,pra,reprint,numerical,amsmath,amssymb,amsfonts,showkeys,showpacs,preprintnumbers]{revtex4-1}
\usepackage{bm,graphicx,xcolor,hyperref,rotating,lineno,multirow,subfigure}

\newcommand{\lam}{\lambda}

\DeclareMathOperator{\Li}{Li}

\begin{document}

\title{Ground state of two electrons on concentric spheres}
\author{Pierre-Fran\c{c}ois Loos}
\email{loos@rsc.anu.edu.au}
\author{Peter M. W. Gill}
\thanks{Corresponding author}
\email{peter.gill@anu.edu.au}
\affiliation{Research School of Chemistry, Australian National University, Canberra, Australian Capital Territory, 0200, Australia}
\date{\today}

\begin{abstract}
We extend our analysis of two electrons on a sphere [Phys. Rev. A {\bf 79}, 062517 (2009); Phys. Rev. Lett. {\bf 103}, 123008 (2009)] to electrons on concentric spheres with different radii.  The strengths and weaknesses of several electronic structure models are analyzed, ranging from the mean-field approximation (restricted and unrestricted Hartree-Fock solutions) to configuration interaction expansion, leading to near-exact wave functions and energies. The M{\o}ller-Plesset energy corrections (up to third-order) and the asymptotic expansion for the large-spheres regime are also considered.  We also study the position intracules derived from approximate and exact wave functions.  We find evidence for the existence of a long-range Coulomb hole in the large-spheres regime, and infer that unrestricted Hartree-Fock theory over-localizes the electrons.
\end{abstract}

\keywords{electron correlation, Hartree-Fock solution, M{\o}ller-Plesset, perturbation theory, 
zero-point oscillations, configuration interaction expansion, position intracule, Coulomb hole.}
\pacs{31.15.ac, 31.15.ve, 31.15.xp, 31.15.xr, 31.15.xt}
\maketitle

\section{Introduction}
In recent work, we have reported near-exact \cite{LoosPRA2009} and exact \cite{LoosPRL2009} solutions of the singlet ground state of two electrons, interacting {\em via} a Coulomb potential, but trapped on the surface of a sphere.  This model was first used by Berry and co-workers in the 1980's to provide insight into angular correlation in two-electron systems \cite{EzraPRA1982, OjhaPRA1987, HindePRA1990}.  It has proven useful for understanding the electronic polarity of nanoclusters and for explaining the giant polarizability of Na$_{14}$F$_{13}$ and spontaneous dipole formation on niobium clusters \cite{ShytovPRB2006}.  Within the adiabatic connection in density functional theory (DFT) \cite{HohenbergPRB1964, KohnPRA1965, ParrYang}, Seidl carefully studied this system \cite{SeidlPRA2007b,SeidlPRA2010} in order to test the ISI (interaction-strength interpolation) model \cite{SeidlPRL2000}, deriving values of the energy by numerical integration.  Furthermore, it has been shown that this kind of spherical constraint applied to the Moshinsky atom \cite{Moshinsky68} leads to a solvable Schr{\"o}dinger equation \cite{LoosPRA2010}.

Berry and collaborators also considered an extension in which each particle is confined to a different, concentric sphere \cite{EzraPRA1983} and used this model to simulate the rovibrational spectra of the water molecule in both the ground \cite{NatansonJCP1984} and excited states \cite{NatansonJCP1986}.  More recently, the model has been applied to quantum-mechanical calculations of large-amplitude light-atom dynamics in polyatomic hydrides \cite{DeskevichJCP2005, DeskevichJCP2008}.

It seems timely therefore to generalize our earlier studies \cite{LoosPRA2009, LoosPRL2009} to the case of two electrons located on the surface of two concentric spheres of different radii.  To be consistent with our previous work \cite{LoosPRA2009,LoosPRL2009},  we will focus on the singlet ground state, which allows us to confine our attention to the symmetric spatial part of the wave function and ignore the spin coordinates.  However, when the two radii are not equal, the spin coordinates are irrelevant, and one can easily generalize the present results to the triplet state by antisymmetrizing the spatial wave function.

Symmetric and asymmetric Hartree-Fock (HF) solutions are discussed in Section \ref{sec:HF} and the strengths and weaknesses of M{\o}ller-Plesset (MP) perturbation theory \cite{MollerPhysRev1934} in Section \ref{sec:MP}.  We consider asymptotic solutions for the large-spheres regime in Section \ref{sec:LargeR} and, in Section \ref{sec:CI}, we study the convergence behavior of the variational configuration interaction (CI) scheme.  Finally, by investigating the shape of the position intracule and the corresponding Coulomb hole (Section \ref{sec:P}), we report the existence of a secondary Coulomb hole, shedding light on long-range correlation effects in two-electron systems.
Atomic units are used throughout.

\section{Hamiltonian}
Our model consists of two concentric spheres of radii $R_1 \le R_2$, each bearing one electron.  The position of the $i$-th electron is defined by the spherical angles $(\theta_i,\phi_i)$, the interelectronic angle $\theta$ by
\begin{equation}\label{cos12}
	\cos \theta = \cos \theta_1 \cos \theta_2 + \sin \theta_1 \sin \theta_2 \cos( \phi_1 - \phi_2 ),
\end{equation}
and the interelectronic distance by $u = | \bm{r}_1 - \bm{r}_2 |$ where
\begin{equation}
	R_2 - R_1 \le u \le R_2 + R_1.
\end{equation}

The Hamiltonian of the system $\Hat{H}$ is simply
\begin{equation}
	\Hat{H} = \Hat{T} + u^{-1},
\end{equation}
where $\Hat{T} = \Hat{T}_1 + \Hat{T}_2 = - (\nabla_1^2 + \nabla_2^2)/2$ is the kinetic energy operator and $u^{-1}$ is the Coulomb operator.  It is sometimes convenient to recast $\Hat{H}$ in term of the interelectronic angle $\theta$.  Introducing the dimensionless parameter $0 < \lam \equiv R_1/R_2 \le 1$ and using \eqref{cos12}, one finds
\begin{equation} \label{H-theta}
	\Hat{H} = - \frac{(1 + \lam^2)}{2 R_1^2 \sin \theta} \frac{d}{d \theta} \left( \sin \theta \frac{d}{d \theta} \right)
		+ \frac{1}{R_2 \sqrt{1 + \lam^2 - 2 \lam \cos \theta}},
\end{equation}
which shows the different scaling behavior of the kinetic and electrostatic terms.

\section{Hartree-Fock approximation} \label{sec:HF}

\begin{figure}
\begin{center}
	\includegraphics[width=0.48\textwidth]{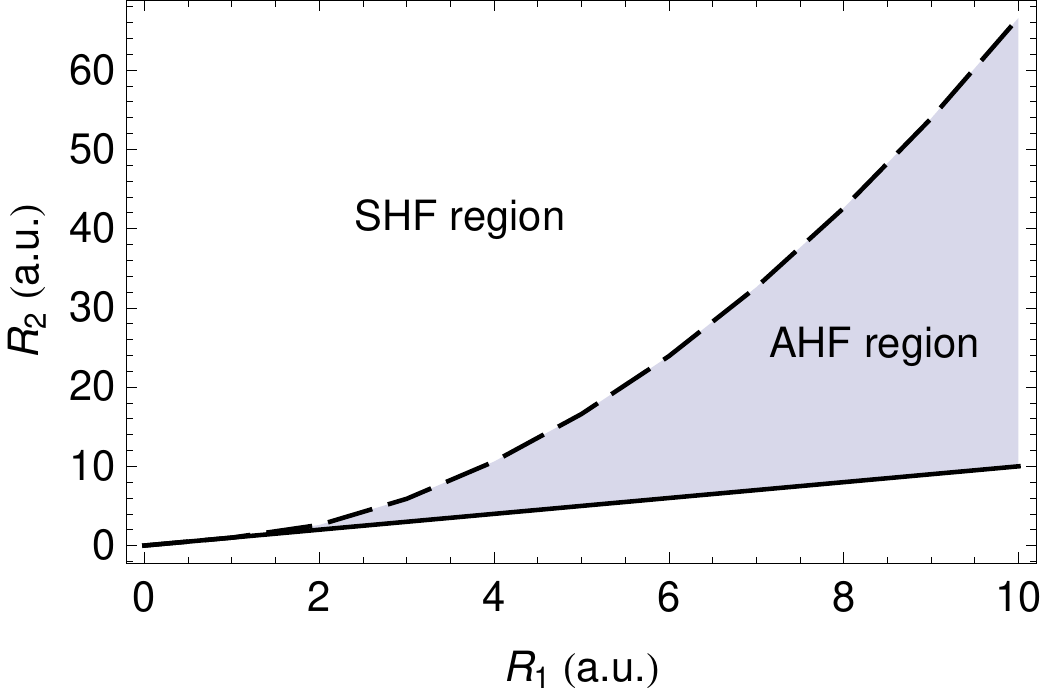}
\caption{\label{fig:SvsA}Frontier between the SHF and AHF solutions with respect to $R_1$ and $R_2$.}
\end{center}
\end{figure}

\begin{figure}
\begin{center}
	\includegraphics[width=0.48\textwidth]{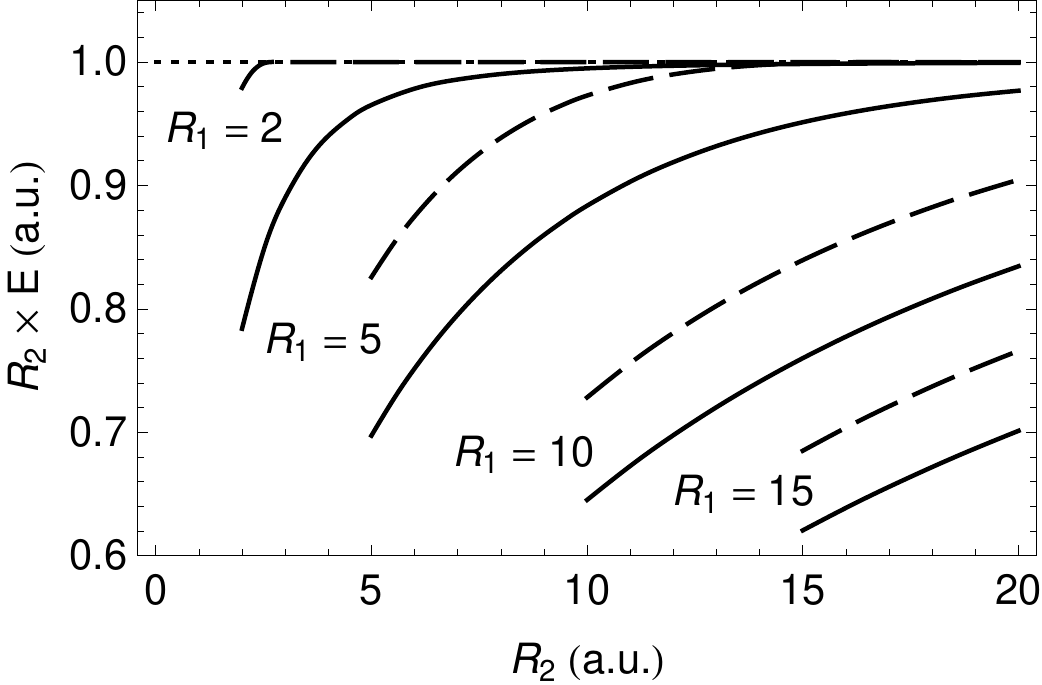}
\caption{\label{fig:AHF}$R_2 \times E^{\rm SHF}$ (dotted), $R_2\times E^{\rm AHF}$ (dashed) and $R_2\times E^{\rm exact}$ (solid) as a function of $R_2$ for $R_1 = 2, 5, 10, 15$.}
\end{center}
\end{figure}

\subsection{Symmetric solution} \label{subsec:SHF}
For $R_1= R_2 = R$, the restricted Hartree-Fock (HF) wave function and energy take \cite{LoosPRA2009} the simple forms
\begin{align}
	\Phi^{\rm HF} & = \frac{1}{4 \pi R^2},			&	E^{\rm HF} & = \frac{1}{R}.
\end{align}

For $R_1 < R_2$, the electrons occupy different orbitals and an unrestricted HF treatment is required.  However, the high symmetry of the system implies that there is a solution in which each orbital is constant over its sphere, and the resulting wave function and energy are
\begin{align}
	\Phi^{\rm SHF} & = \frac{1}{4 \pi R_1 R_2},	&	E^{\rm SHF} & = \frac{1}{R_2}.
\end{align}
We call this the ``symmetric Hartree-Fock'' (SHF) solution, because the orbitals are spherically symmetric and we note that the SHF energy depends only on the radius of the larger sphere.

\subsection{Asymmetric solution} \label{subsec:AHF}

For certain values of $R_1$ and $R_2$, a second, lower-energy HF solution arises \cite{CizekJCP1967, CizekJCP1970, SeegerJCP1977}, in which the two electrons tend to localize on opposite sides of the spheres.  We call this the ``asymmetric Hartree-Fock'' (AHF) solution for the orbitals possess cylindrical, not spherical, symmetry.

To obtain the AHF wave function
\begin{equation}
	\Phi^{\rm AHF} (\theta_1,\theta_2) = \Psi_1 (\theta_1)\,\Psi_2 (\theta_2),
\end{equation}
we expand the orbitals as
\begin{align} \label{basis-AHF}
	\Psi_1 (\theta_1) & = \sum_{\ell=0}^\infty c_\ell \,\Psi_\ell (\theta_1),	&
	\Psi_2 (\theta_2) & = \sum_{\ell=0}^\infty d_\ell \,\Psi_\ell (\theta_2),
\end{align}
in the basis of zonal harmonics \cite{Abramowitz}
\begin{equation}
	\Psi_\ell (\theta_i)= Y_\ell (\theta_i)/R_i \equiv Y_{\ell 0} (\theta_i,\phi_i)/R_i.
\end{equation}

The Fock matrix elements for the two orbitals are
\begin{gather}
	F_1^{\ell_1 \ell_2} = \frac{\ell_1(\ell_1+1)}{2 R_1^2} \delta_{\ell_1,\ell_2}
						+ \sum_{\ell_3,\ell_4=0}^{\infty} d_{\ell_3} d_{\ell_4} G_{\ell_1 \ell_2}^{\ell_3 \ell_4},	\\
	F_2^{\ell_1 \ell_2} = \frac{\ell_1(\ell_1+1)}{2 R_2^2} \delta_{\ell_1,\ell_2}
						+ \sum_{\ell_3,\ell_4=0}^{\infty} c_{\ell_3} c_{\ell_4} G_{\ell_1 \ell_2}^{\ell_3 \ell_4},
\end{gather}
where $\delta_{\ell_1,\ell_2}$ is the Kronecker symbol and 
\begin{equation} \label{G}
	G_{\ell_1 \ell_2}^{\ell_3 \ell_4} = \frac{(-1)^{\ell_3+\ell_4}}{R_2} \sum_\ell \frac{4 \pi}{2\ell+1} \lam^\ell
										\left< \ell_1\,\ell_2\,\ell \right> \left< \ell_3\,\ell_4\,\ell \right>
\end{equation}
are the two-electron integrals expressed in terms of the Wigner $3j$-symbols \cite{Edmonds}
\begin{equation} \label{Y-W3J}
	\left< \ell_1 \ell_2 \ell_3 \right> = \sqrt{\frac{(2\ell_1+1)(2\ell_2+1)(2\ell_3+1)}{4 \pi}}
												\begin{pmatrix}		\ell_1	&	\ell_2	&	\ell_3	\\
																		0	&		0	&		0	\\
												\end{pmatrix}^2.
\end{equation}
The summation in \eqref{G} runs from $\max (|\ell_1 - \ell_2|,|\ell_3 - \ell_4|)$ to $\min (\ell_1 + \ell_2,\ell_3 + \ell_4)$ 
because of selection rules \cite{Edmonds}.

The AHF energy is
\begin{multline}\label{E-AHF}
	E^{\rm AHF} = \frac{1}{2} \sum_{\ell=0}^\infty \left[ c_\ell^2\ \frac{\ell(\ell+1)}{2R_1^2} + d_\ell^2\ \frac{\ell(\ell+1)}{2R_2^2} \right]\\
	+ \frac{1}{2} \sum_{\ell_1,\ell_2=0}^\infty \left( c_{\ell_1} c_{\ell_2} F_1^{\ell_1 \ell_2} + d_{\ell_1} d_{\ell_2} F_2^{\ell_1 \ell_2} \right).
\end{multline}
For all of the radii considered, truncating the expansions in \eqref{basis-AHF} at $\ell = 15$ yields $E^{\rm AHF}$ with an accuracy of $10^{-12}$.

The asymptotic limits of the HF energies satisfy
\begin{gather}
	\lim_{R_2 \to \infty} R_2\,E^{\rm SHF} = 1,	\\
	\lim_{R_1,R_2 \to \infty} (R_1 + R_2)\,E^{\rm AHF} = 1,
\end{gather}
and the limiting AHF energy corresponds to the Coulomb interaction between two electrons that are fully localized on opposite side of their respective spheres.  Such systems are known as Wigner molecules \cite{WignerPR1934} and have been observed in a variety of similar systems \cite{AlaviJCP2000, ThompsonPRB2002, ThompsonPRB2004, ThompsonJCP2005}.

The creation of localized orbitals leads to decreased Coulombic repulsion but increased kinetic energy, and an asymmetric solution therefore exists only when the former outweighs the latter.  By considering an orbital basis consisting of only $Y_0$ and $Y_1$, it can be shown that this occurs only when $R_1 > R^{\rm crit}$ and $R_2 < R_1^2/R^{\rm crit}$, where $R^{\rm crit} = 3/2$.  Figure \ref{fig:SvsA} illustrates this graphically.

Figure \ref{fig:AHF} shows the SHF, AHF and exact energies as functions of $R_2$ for several values of $R_1$.   The difference $E^{\rm AHF} - E^{\rm exact}$ decreases as $R_1 = R_2$ increases, indicating that the AHF energy is asymptotically correct.

\section{Expansion for Small Spheres} \label{sec:MP}

\begin{figure}
	\begin{center}
	\includegraphics[width=0.48\textwidth]{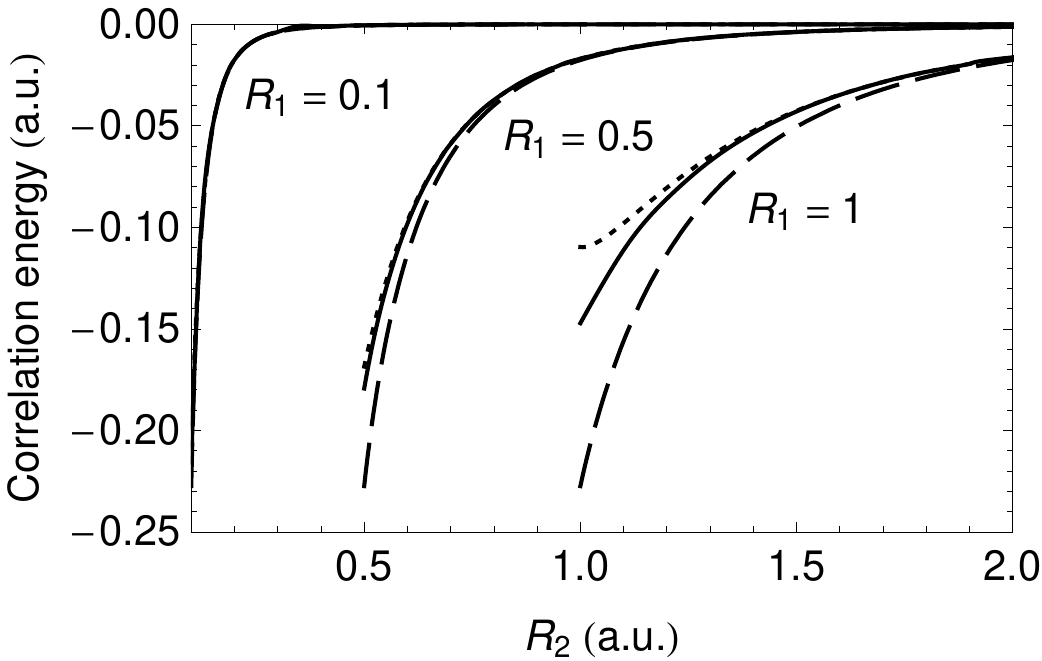}
	\caption{\label{fig:MPn} MP2 (dashed), MP3 (dotted) and exact (solid) correlation energies (relative to SHF) as a function of $R_2$
for $R_1 = 0.1, 0.5, 1$.}
	\end{center}
\end{figure}

\subsection{First-order wave function}
In M{\o}ller-Plesset (MP) perturbation theory \cite{MollerPhysRev1934}, the total Hamiltonian is partitioned into a zeroth-order Hamiltonian $\Hat{H}_0 = \Hat{T}$ and a perturbative correction $\Hat{V} = u^{-1}$.  The unperturbed orbitals are spherical harmonics on each sphere and therefore, from Section \ref{subsec:SHF}, we have $E^{(0)} = 0$ and $E^{(1)} = R_2^{-1}$.

The $\ell$-th excited eigenfunction and eigenvalue of $\Hat{H}_0$ with $S$ symmetry are \cite{Slater, Edmonds, LoosPRA2009}
\begin{gather}
	\Phi_\ell(\theta) = \frac{\sqrt{2\ell+1}}{4 \pi R_1 R_2} P_\ell (\cos\theta),	\\
	E_\ell = \ell(\ell+1)\left(\frac{1}{2R_1^2}+\frac{1}{2R_2^2}\right).
\end{gather}
In intermediate normalization, the first-order correction to the wave function is
\begin{align} \label{Phi-1}
	\Phi^{(1)}	& = \sum_{\ell=1}^\infty \frac{\left< \Phi_0 \left| u^{-1} \right| \Phi_l \right>}{E_0 - E_\ell} \Phi_\ell(\theta)	\notag	\\
				& = - \frac{1}{2\pi R_1} \frac{\lam^2}{1+\lam^2}\ Z(\cos\theta,\lam),
\end{align}
where it can be shown that
\begin{align}
	Z(x,\lam)	& = \sum_{\ell=1}^\infty \frac{\lam^\ell}{\ell(\ell+1)} P_\ell(x)		\notag	\\
				& = 1 + \log 2 - \log(1 - \lam x + \sqrt{1 - 2\lam x + \lam^2})		\notag	\\
				& + \frac{\log(1-x) - \log(\lam - x + \sqrt{1 - 2\lam x + \lam^2})}{\lam}.
\end{align}
This yields the normalized first-order wave function
\begin{equation}
	\Phi^{\rm MP1}(u) = \frac{\Phi_0 + \Phi^{(1)}}
	{\sqrt{1+\left[\frac{2\lam R_1}{1+\lam^2}\right]^2 \sum_{\ell=1}^\infty \frac{\lam^{2\ell}}{\ell^2(\ell+1)^2(2\ell+1)}}}.
\end{equation}

\subsection{Second-order energy}
Using \eqref{Phi-1}, one finds that the second-order energy
\begin{align}
	E^{(2)}	& = \left< \Phi_0 \left| u^{-1} \right| \Phi^{(1)} \right>									\notag	\\
			& = \frac{2(1-\lam)^2 \ln(1-\lam) + 2(1+\lam)^2 \ln(1+\lam) - 6\lam^2}{1+\lam^2}	\notag	\\
			& = -\frac{\lam^4}{1+\lam^2}\ \frac{{_2F_1}(1,2,5,\lam)+{_2F_1}(1,2,5,-\lam)}{6}	\notag	\\
			& = - \frac{\lam^4}{3} + \frac{4 \lam^6}{15} + O(\lam^8),
\end{align}
(where $_2F_1$ is the Gauss hypergeometric function \cite{Abramowitz}) depends only on the ratio of the radii.

When the radii are equal, $E^{(2)}$ takes the value
\begin{equation}
	\lim_{\lam \to 1} E^{(2)} = 4 \ln 2 - 3,
\end{equation}
which has been discussed by Seidl \cite{SeidlPRA2007b, SeidlPRA2010} and us \cite{LoosPRA2009, LoosJCP2009}.  When the radii are very different (\textit{i.e.}~$\lam \approx 0$), the HF treatment is accurate and the second-order energy is
\begin{equation} \label{eq:Edisp}
	E^{(2)} \sim C_4 / R_2^4,
\end{equation}
where $C_4 = - R_1^4 / 3$.  Although \eqref{eq:Edisp} can be identified as the dispersion energy, it does not exhibit the usual $R^{-6}$ behavior.  Analogous results have also been reported for other systems \cite{Dobson06}.

\subsection{Third-order energy}
Using \eqref{Phi-1}, one finds the third-order energy
\begin{align} \label{EMP3}
	E^{(3)}	& = \left< \Phi^{(1)} \left| u^{-1} - E^{(1)} \right| \Phi^{(1)} \right>							\notag	\\
			& = \frac{4\lam R_2}{(1+\lam^2)^2} \biggl[24\lam^3 + 2\lam(1-\lam^2) \Li_2(\lam^2)	\notag	\\
			& \quad - 10\lam(1-\lam)^2 \ln(1-\lam) - (1-\lam)^3 \ln^2(1-\lam)						\notag	\\
			& \quad - 10\lam(1+\lam)^2 \ln(1+\lam) + (1+\lam)^3 \ln^2(1+\lam)  \biggr]				\notag	\\
			& = R_2 \left[\frac{2\lam^8}{9} - \frac{12\lam^{10}}{35} + O(\lam^{12}) \right],
\end{align}
where $\Li_2$ is the dilogarithm function \cite{Lewin}.

When the radii are equal, $E^{(3)}$ takes the value
\begin{equation}
	\lim_{\lam \to 1} E^{(3)} = 8R_2 (3 - 5 \ln 2 + \ln^2 2)
\end{equation}
that we have given previously \cite{LoosPRA2009}.  When the radii are very different and $R_2$ is not too large, $E^{(3)}$ is tiny and $E^{(2)}$ is a good approximation to the total correlation energy.

The MP2 and MP3 correlation energies, defined by
\begin{equation}
        E^{\text{MP}n} = \sum_{m=2}^n E^{(m)},
\end{equation}
are shown in Fig. \ref{fig:MPn}.  For $R_1 = 0.1$, the MP2 and MP3 energies are accurate for all $R_2$.  For larger $R_1$, the discrepancy between the MP and exact energies is noticeable for small $R_2$, but remains small for large $R_2$.

The MP3 energy is usually better than the MP2 energy.  However, as we have shown previously \cite{LoosPRA2009}, the MP expansion appears to diverge when the radii are similar and not small \cite{SeidlPRL2000, SeidlPRA2007b}.

\section{Expansion for Large Spheres} \label{sec:LargeR}

\begin{figure}
	\begin{center}
	\includegraphics[width=0.48\textwidth]{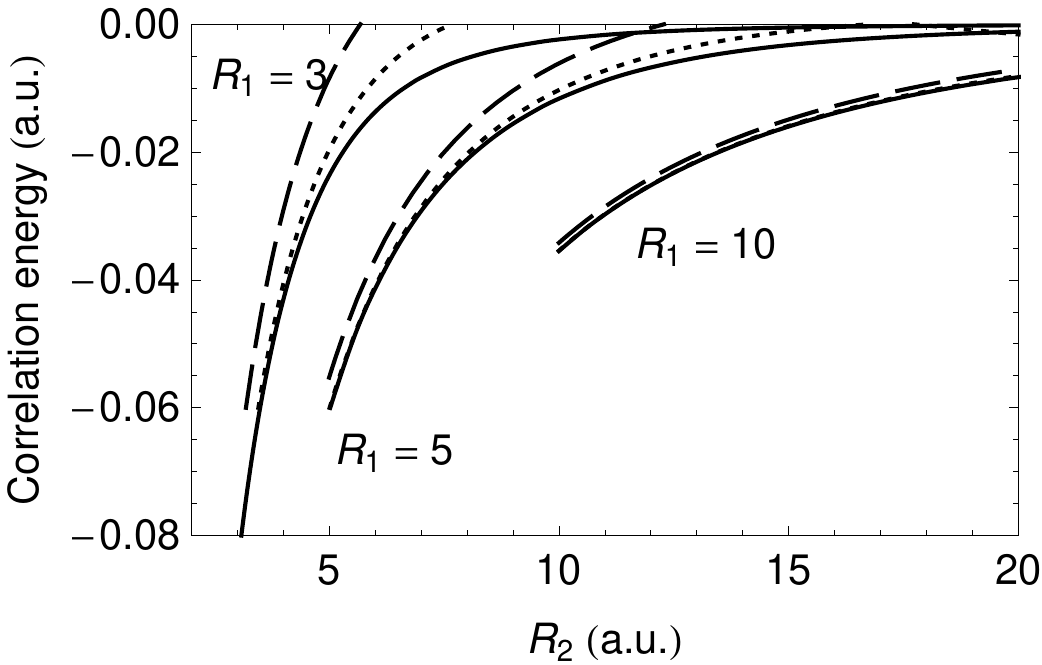}
	\caption{\label{fig:LargeR}$E^{\rm LS0}$ (dashed) and $E^{\rm LS1}$ (dotted) correlation energies (relative to SHF) as a function of
$R_2$ for $R_1 = 3, 5, 10$.  The exact correlation energy (solid) is also shown.}
	\end{center}
\end{figure}

\subsection{Harmonic approximation}

In the large-spheres (LS) regime, the electrons reduce their Coulomb repulsion by localizing on opposite sides of their spheres, oscillating around their equilibrium positions with angular frequency $\omega$ (zero-point oscillations).  The same phenomenon has been observed by Seidl and collaborators \cite{SeidlPRA1999a, SeidlPRA1999b, SeidlPRL2000, SeidlPRA2007a, SeidlPRA2007b, GoriGiorgiJCTC2009, GoriGiorgiPRL2009}.

In this case, the supplementary angle $\xi = \pi - \theta$ becomes the natural coordinate of the system.  Using the Taylor expansions, $\cot \xi = \frac{1}{\xi} + O(\xi)$ and 
\begin{equation}\label{Coulomb-Taylor}
	\frac{1}{\sqrt{1 + \lam^2 + 2 \lam \cos \xi}} = \frac{1}{(1 + \lam)} + \frac{\lam}{2 (1 + \lam)^3} \xi^2 + O(\xi^4),
\end{equation}
the Hamiltonian \eqref{H-theta} becomes
\begin{multline} \label{H-xi}
	\Hat{H}^{\omega} = - \frac{1 + \lam^2}{2 R_1^2} \left( \frac{d^2}{d \xi^2} + \xi \frac{d}{d \xi} \right)\\
						+ \frac{\lam}{(1 + \lam)R_1} \left[ 1 + \frac{\lam\,\xi^2}{2\,(1 + \lam)^2} \right].
\end{multline}
The lowest eigenfunction of \eqref{H-xi} is
\begin{equation}\label{Phi-LR}
	\Phi^{\omega} (\xi) \propto \exp \left[ - \frac{1}{2} \sqrt{\frac{\lam^2 R_1}{(1 + \lam)^3(1 + \lam^2)}} \xi^2\right],
\end{equation}
and the associated eigenvalue is
\begin{equation}\label{E-Upsilon}
	 E^{\rm LS0} = E^{e-e} + E^{\omega} = \frac{1}{R_1 + R_2} + \frac{\omega}{2}.
\end{equation}
The first term of \eqref{E-Upsilon} represents the classical interaction of two electrons separated by a distance $R_1 + R_2$,
and the second one is the energy associated with the zero-point oscillations of angular frequency 
\begin{equation}
	\omega = \frac{2\sqrt{(1 + \lam^2)/\lam}}{(R_1 + R_2)^{3/2}}.
\end{equation}

\subsection{First anharmonic correction}
The first anharmonic correction
\begin{equation}
	\Hat{W} = \frac{(1 + \lam^2)}{6 R_1^2} \xi \frac{d}{d\xi} - \frac{\lam^2(\lam^2 - 7 \lam + 1)}{24 (1+\lam)^5 R_1} \xi^4
\end{equation}
arises from the next two terms of the Taylor expansion of $\cot \xi$ and the Coulomb operator \eqref{Coulomb-Taylor}.  Defining $E^{\rm LS1} = E^{\rm LS0} + \mathcal{E}^{(1)}$, the anharmonic correction energy is
\begin{align}
	\mathcal{E}^{(1)}	& = 4 \pi^2 R_1^2 R_2^2 \int_0^{\infty} \Phi^{\omega} (\xi)\,\Hat{W}\,\Phi^{\omega} (\xi) \xi d\xi	\notag	\\
						& = - \frac{(1 - \lam + \lam^2)(1 + \lam^2)}{4(R_1 + R_2)^2}.
\end{align}
The LS0 and LS1 correlation energies are shown in Fig. \ref{fig:LargeR} with respect to $R_2$ for three values of $R_1$.  For the large values of $R_1$, both curves agree very well with the exact correlation energies, while for the smaller values of the 
radius of the first sphere, LS1 systematically improves the results compared to LS0.

\section{Configuration interaction} \label{sec:CI}

\begin{table*}
\caption{\label{tab:CI} Convergence of correlation energy with respect to the number $L$ of terms in the CI wave function.}
\begin{ruledtabular}
\begin{tabular}{ccccc}
	$L$	& $R_1 = 1 \quad R_2 = 1$	& $R_1 = 1 \quad R_2 = 1.1$	&$R_1 = 1 \quad R_2 = 1.5$	& $R_1 = 1 \quad R_2 = 2$	\\
	\hline
	1	&$-0.131\ 665\ 623\ 696$		&$-0.102\ 135\ 552\ 400$		&$-0.041\ 049\ 324\ 810$		&$-0.015\ 832\ 811\ 848$		\\
	2	&$-0.141\ 241\ 198\ 782$		&$-0.108\ 514\ 851\ 797$		&$-0.042\ 647\ 613\ 578$		&$-0.016\ 238\ 558\ 022$		\\
	3	&$-0.144\ 065\ 402\ 167$		&$-0.110\ 102\ 786\ 034$		&$-0.042\ 870\ 984\ 637$		&$-0.016\ 271\ 251\ 778$		\\
	4	&$-0.145\ 273\ 783\ 726$		&$-0.110\ 674\ 429\ 216$		&$-0.042\ 915\ 528\ 628$		&$-0.016\ 274\ 960\ 819$		\\
	5	&$-0.145\ 900\ 461\ 200$		&$-0.110\ 923\ 205\ 726$		&$-0.042\ 926\ 179\ 125$		&$-0.016\ 275\ 462\ 730$		\\
	10	&$-0.146\ 847\ 645\ 782$		&$-0.111\ 180\ 386\ 287$		&$-0.042\ 930\ 208\ 158$		&$-0.016\ 275\ 553\ 424$		\\
	15	&$-0.147\ 047\ 095\ 403$		&$-0.111\ 201\ 118\ 465$		&$-0.042\ 930\ 221\ 857$		&$-0.016\ 275\ 553\ 441$		\\
	20	&$-0.147\ 120\ 296\ 106$		&$-0.111\ 204\ 056\ 487$		&$-0.042\ 930\ 221\ 942$		&$-0.016\ 275\ 553\ 441$		\\
	25	&$-0.147\ 155\ 035\ 738$		&$-0.111\ 204\ 595\ 556$		&$-0.042\ 930\ 221\ 942$		&$-0.016\ 275\ 553\ 441$		\\
	30	&$-0.147\ 174\ 201\ 368$		&$-0.111\ 204\ 710\ 528$		&$-0.042\ 930\ 221\ 942$		&$-0.016\ 275\ 553\ 441$		\\
	35	&$-0.147\ 185\ 880\ 267$		&$-0.111\ 204\ 737\ 604$		&$-0.042\ 930\ 221\ 942$		&$-0.016\ 275\ 553\ 441$		\\
	40	&$-0.147\ 193\ 518\ 573$		&$-0.111\ 204\ 744\ 445$		&$-0.042\ 930\ 221\ 942$		&$-0.016\ 275\ 553\ 441$		\\
	45	&$-0.147\ 198\ 785\ 870$		&$-0.111\ 204\ 746\ 267$		&$-0.042\ 930\ 221\ 942$		&$-0.016\ 275\ 553\ 441$		\\
	50	&$-0.147\ 202\ 570\ 742$		&$-0.111\ 204\ 746\ 773$		&$-0.042\ 930\ 221\ 942$		&$-0.016\ 275\ 553\ 441$		\\
	\hline
	Exact	&$-0.147\ 218\ 934\ 944$	&$-0.111\ 204\ 746\ 979$		&$-0.042\ 930\ 221\ 942$		&$-0.016\ 275\ 553\ 441$		\\
\end{tabular}
\end{ruledtabular}
\end{table*}

To obtain an accurate wave function, we expand it in the Legendre basis
\begin{equation}\label{Psi-CI}
	\Phi_L^{\rm CI} (\theta) = \sum_{\ell=0}^{L} T_{\ell}\,\Phi_{\ell} (\theta),
\end{equation}
where $T_{\ell}$ is the CI amplitude of the excited configuration $\Phi_{\ell}$.  The elements of the CI matrix are given by
\begin{equation} \label{H-CI}
	\Bigl< \Phi_{\ell_1} \Bigl| \Hat{H} \Bigr| \Phi_{\ell_2} \Bigr>
	= \left[ \frac{\ell_1(\ell_1+1)}{2 R_1^2} + \frac{\ell_2(\ell_2+1)}{2 R_2^2} \right] \delta_{\ell_1,\ell_2}
	+ \frac{1}{R_2} \sum_{\ell=|\ell_1-\ell_2|}^{\ell_1+\ell_2} \sqrt{\frac{4\pi}{2\ell+1}} \lam^\ell \left< \ell_1\,\ell_2\,\ell \right>,
\end{equation}
where $\left< \ell_1\,\ell_2\,\ell\right>$ is given by \eqref{Y-W3J}. 

In our earlier work on the $R_1 = R_2$ case \cite{LoosPRA2009}, we found that the CI expansion converges slowly with respect to $L$ because of the interelectronic cusp that arises wherever the electrons meet \cite{Kato1957}.  We also showed that this problem can be overcome by expanding the wave function as a polynomial in $u$.

Here, however, we find (Table \ref{tab:CI}) that the CI expansion converges rapidly, provided that $R_2$ is significantly greater than $R_1$.  This is to be expected, because the fact that the electrons are confined to different spheres means that they can never meet and that the exact wave function is therefore cuspless.

\section{Intracules and Holes} \label{sec:P}

To study the \emph{relative} positions of the electrons in space, we have computed the position intracule
\begin{equation}
	\mathcal{P}(u) = \left< \Phi | \delta(|\bm{r}_1 - \bm{r}_2 | - u) | \Phi \right>
\end{equation}
the probability density for the interelectronic separation $u$, from several of the wave functions $\Phi$ above.  Because the SHF, MP1, LR and CI wave functions depend only on $u$ (or, equivalently, on the interelectronic angle), their position intracules are given by the simple Jacobian-weighted density
\begin{equation}
	\mathcal{P}(u) = 8 \pi^2 R_1 R_2\,u\,\left| \Phi (u) \right|^2.
\end{equation}
For $R_1 = R_2$, the MP2 intracule is also available \cite{LoosPRA2009}.

The SHF intracule 
\begin{equation}
	\mathcal{P}^{\rm SHF}(u) = \begin{cases}
									\frac{u}{2 R_1 R_2},		&	R_2 - R_1 \le u \le R_1 + R_2,	\\
										\quad 0,				&	\text{otherwise},
								\end{cases}
\end{equation}
grows linearly over the domain of allowed $u$ values.  The AHF intracule is more complicated but is given by
\begin{equation}
	\mathcal{P}^{\rm AHF}(u) = \mathcal{P}^{\rm SHF}(u) \sum_{\ell_1=0}^\infty \Tilde{c}_{\ell_1}P_{\ell_1}(x) \sum_{\ell_2=0}^\infty \Tilde{d}_{\ell_2} P_{\ell_2}(x),
\end{equation}
with $\Tilde{c}_{\ell} = \sqrt{2\ell+1}\,c_{\ell}$ and $x = (R_1^2 +R_2^2 - u^2)/(2 R_1 R_2)$.

We define the Coulomb hole \cite{CoulsonProcPhysSoc1961}
\begin{equation}
        \Delta \mathcal{P}(u) = \mathcal{P}(u) - \mathcal{P}^{\rm HF}(u)
\end{equation} 
as the difference between the intracule from a correlated wave function and that from the lowest HF wave function.  To quantify the correlation effects, it is useful to identify the minimum $\Check{u}^{\rm sr}$, the root $\Bar{u}^{\rm sr}$ and the strength
\begin{equation}
	S^{\rm sr} = \int_0^{\Bar{u}^{\rm sr}} | \Delta \mathcal{P}(u) | \,du
\end{equation}
of the short-range (sr) Coulomb hole. In certain cases, a secondary long-range (lr) Coulomb hole appears \cite{PearsonMolPhys2009}.  Its strength is given by
\begin{equation}
	S^{\rm lr} = \int_{\Bar{u}^{\rm lr}}^\infty | \Delta \mathcal{P}(u) | \,du,
\end{equation}
where $\Bar{u}^{\rm lr}$ is the long-range root.

\subsection{Weakly correlated regime}

\begin{figure*}
	\begin{center}
	\subfigure[~$R_1 = R_2 = 0.5$]	{\includegraphics[width=0.48\textwidth]{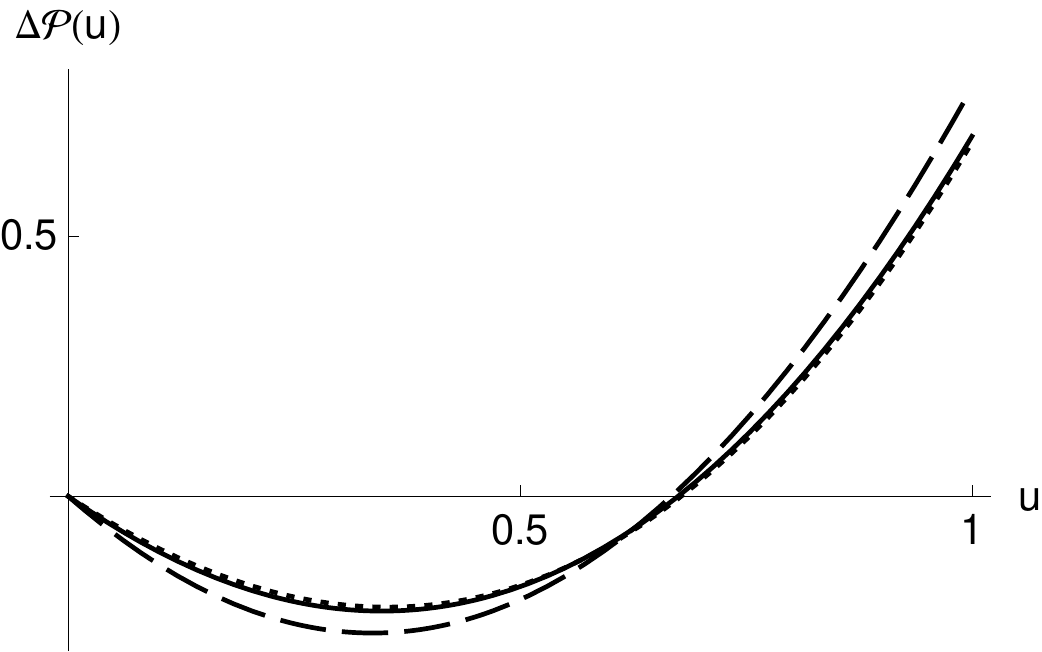}}
	\subfigure[~$R_1 = 0.5$]	{\includegraphics[width=0.48\textwidth]{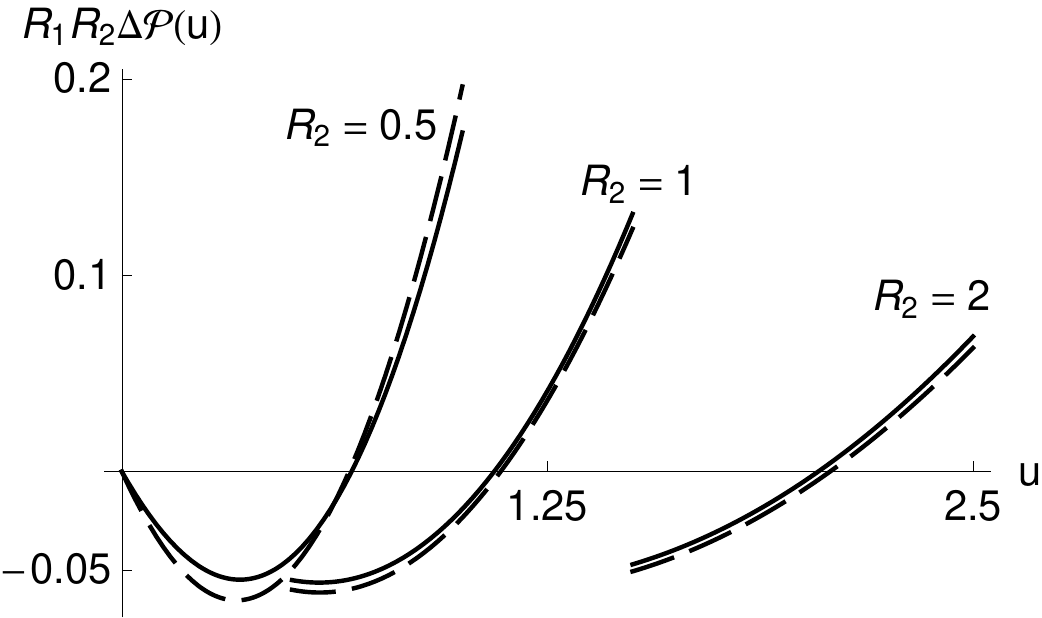}}
	\subfigure[~$R_1 = R_2 = 1$]	{\includegraphics[width=0.48\textwidth]{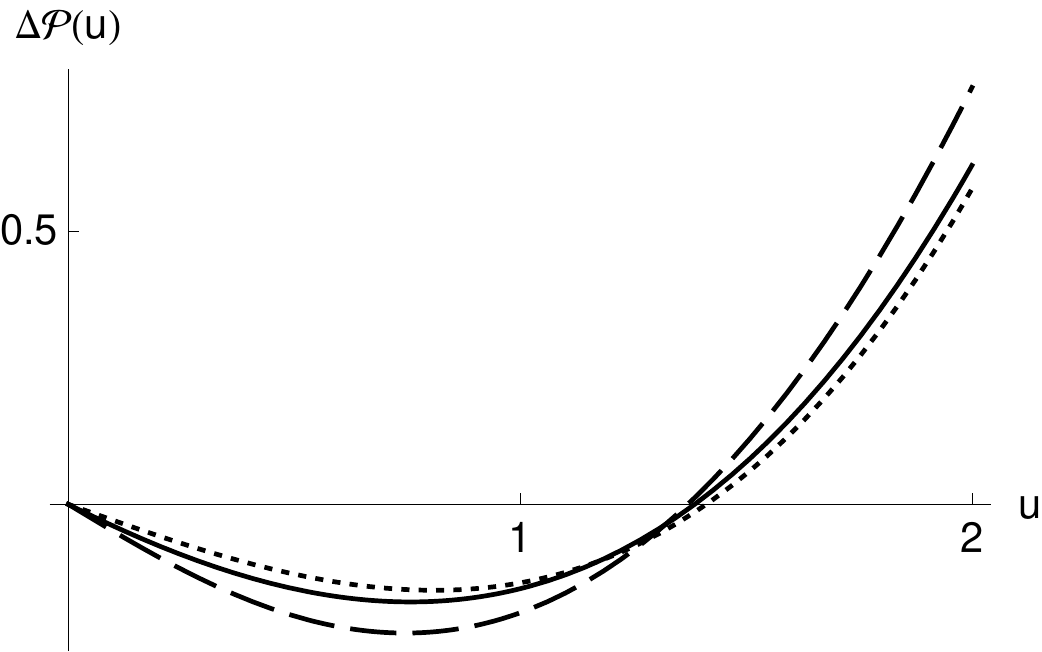}}
	\subfigure[~$R_1 = 1$]		{\includegraphics[width=0.48\textwidth]{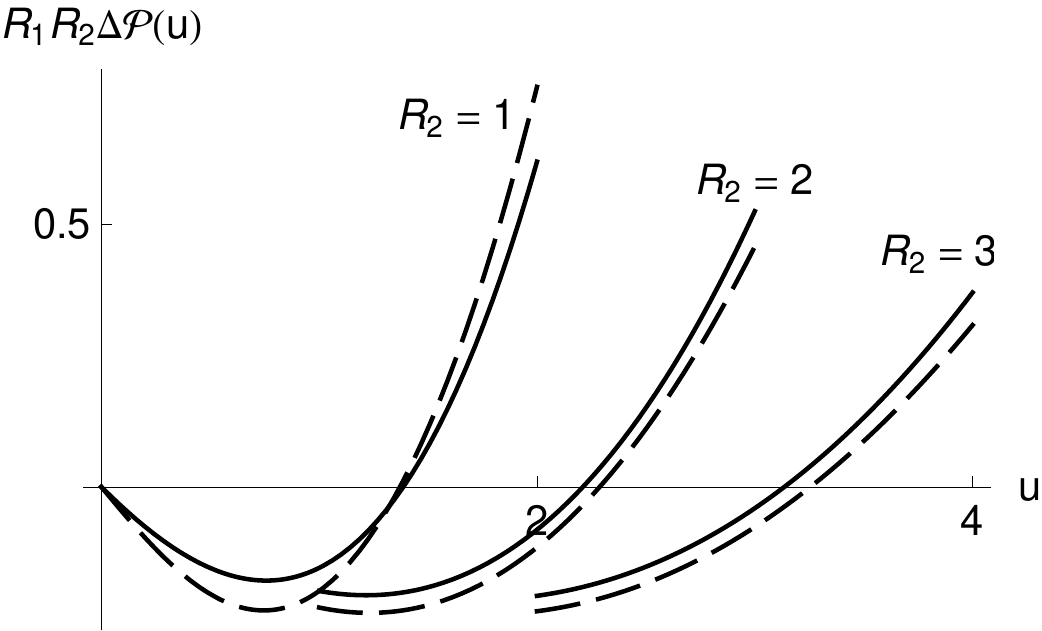}}
	\subfigure[~$R_1 = R_2 = 1.5$]	{\includegraphics[width=0.48\textwidth]{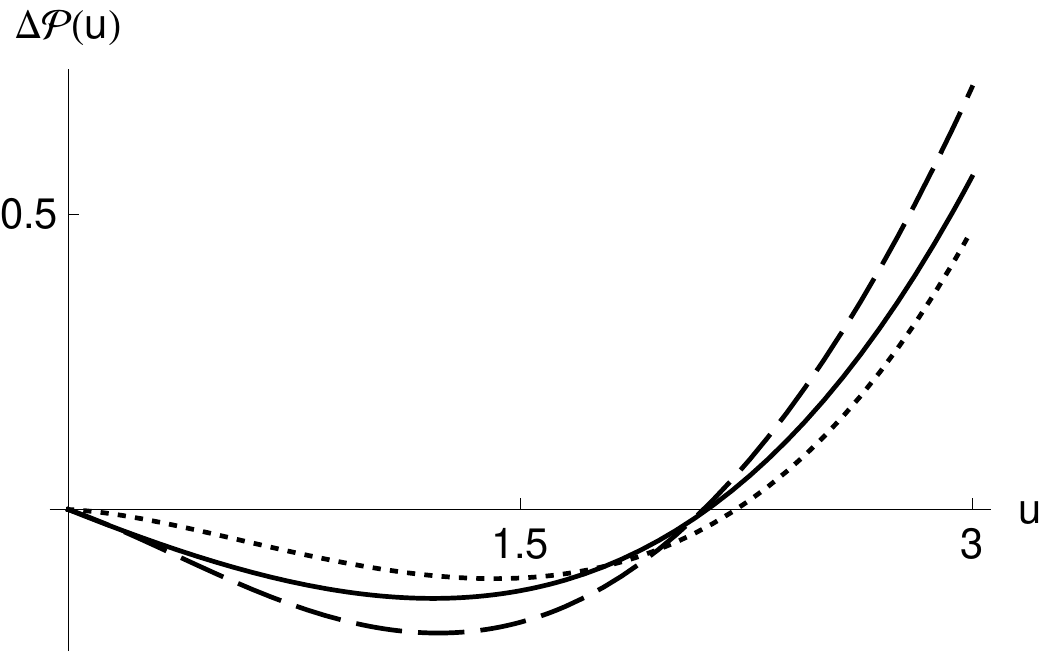}}
	\subfigure[~$R_1 = 1.5$]	{\includegraphics[width=0.48\textwidth]{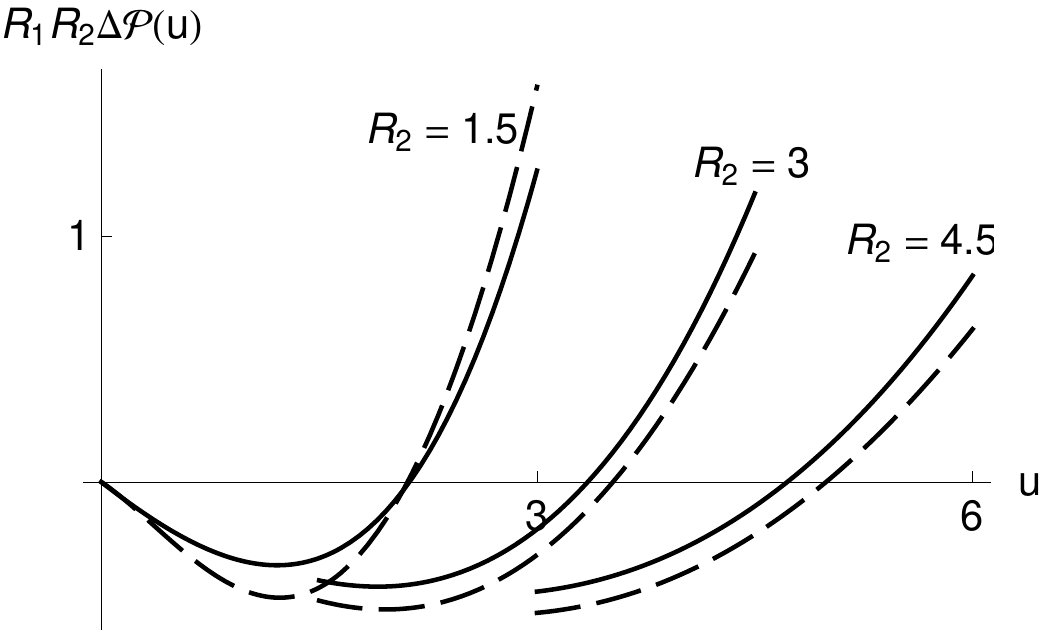}}
	\caption{\label{fig:P-MP-R} On the left: MP1 (dashed), MP2 (dotted) and exact (solid) holes for various $R_1 = R_2$.  On the right: MP1 (dashed) and exact (solid) holes for various $R_1 \le R_2$.}
	\end{center}
\end{figure*}

\begin{figure}
	\begin{center}
	\includegraphics[width=0.48\textwidth]{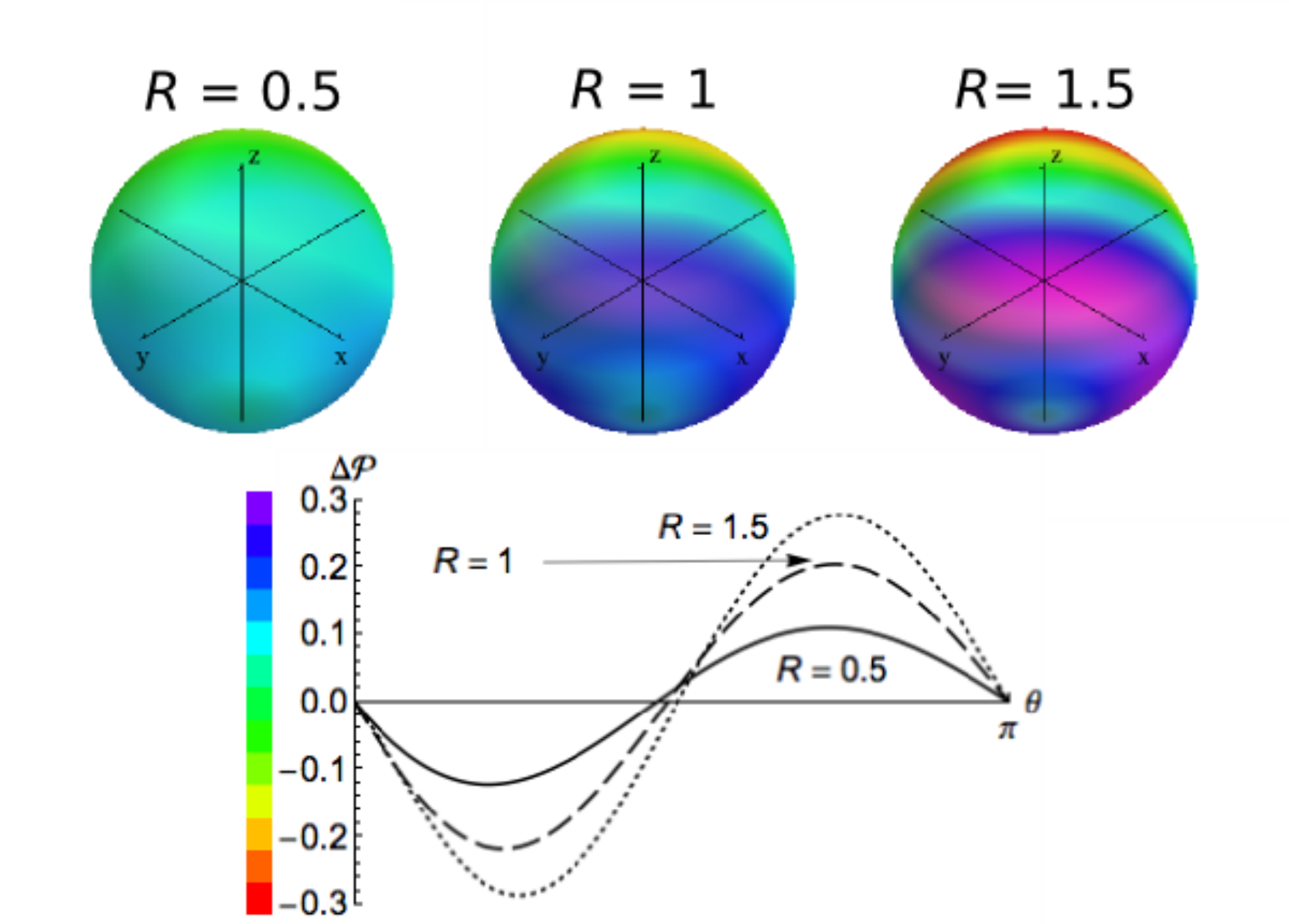}
	\caption{\label{fig:densityplot-MP1}MP1 holes for various $R = R_1 = R_2$ plotted on the surface of a sphere.  The holes with respect to $\theta$ are also represented.}
	\end{center}
\end{figure}

\begin{figure}
	\begin{center}
	\includegraphics[width=0.48\textwidth]{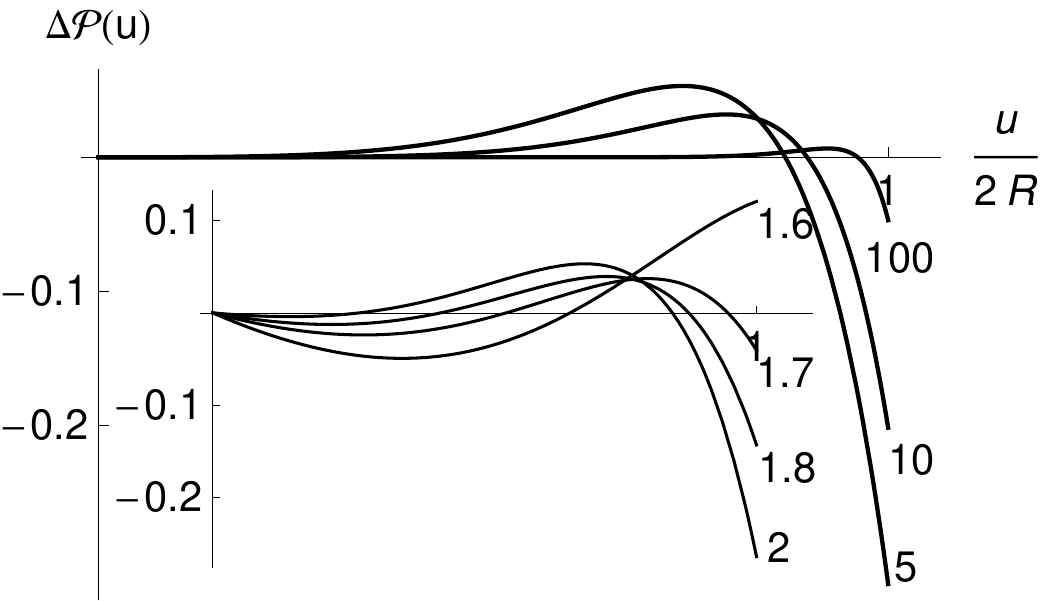}
	\caption{\label{fig:DP-wrt-R} Holes based on the exact wave function for various ($R = 5, 10, 100$).  The inset graph shows holes for radii just above the critical value $R^{\rm crit} = 3/2$.}
	\end{center}
\end{figure}

\begin{table}
\caption{\label{tab:DP-MP} Minimum ($\Check{u}$), root ($\Bar{u}$) and strength ($S$) of the MP1, MP2 and exact Coulomb holes for $R_1 = R_2 = R$.}
\begin{ruledtabular}
\begin{tabular}{cccc}
	$R$	& \multicolumn{3}{c}{Minimum $\Check{u}^{\rm sr}$}\\		
						\cline{2-4} 					
			&	MP1 		&	MP2 		&	Exact		\\
	\hline					                
	0.1		&	0.062 	 	&	0.063		&	0.063		\\
	0.2		&	0.127		&	0.130		&	0.129		\\
	0.5		&	0.337		&	0.353 		&	0.347		\\
	1.0		&	0.746		&	0.810		&	0.757		\\
	1.5		&	1.228		&	1.413		&	1.211		\\
	\hline
			& \multicolumn{3}{c}{Root $\Bar{u}^{\rm sr}$}	\\ 		
						\cline{2-4} 					
			&	MP1		&	MP2		&	Exact		\\
	\hline
	0.1		&	0.130		&	0.131		&	0.131		\\	
	0.2		&	0.262		&	0.264		&	0.264		\\	
	0.5		&	0.667		&	0.678		&	0.675		\\	
	1.0		&	1.371		&	1.412		&	1.386		\\	
	1.5		&	2.109		&	2.216		&	2.121		\\	
	\hline
			& \multicolumn{3}{c}{Strength $S^{\rm sr}$}	\\
						\cline{2-4}
			&	MP1		&	MP2		&	Exact		\\
	\hline                
	0.1		&	0.0245		&	0.0235		&	0.0235		\\
	0.2		&	0.048		&	0.045		&	0.045		\\
	0.5		&	0.116		&	0.096		&	0.099		\\
	1.0		&	0.211		&	0.145		&	0.164		\\
	1.5		&	0.476		&	0.235		&	0.210		\\
\end{tabular}
\end{ruledtabular}
\end{table}

\begin{table}
\caption{\label{tab:DP-LR}Minimum ($\Check{u}$), maximum ($\Hat{u}$), root ($\Bar{u}$) and strength ($S$) of the exact Coulomb hole for various $R$.}
\begin{ruledtabular}
\begin{tabular}{cccc}
	$R$	&					\multicolumn{3}{c}{Short-range Coulomb hole}							\\
												\cline{2-4}
			& Minimum $\Check{u}^{\rm sr}$	&	Root $\Bar{u}^{\rm sr}$	&	Strength $S^{\rm sr}$	\\
	\hline
	1.6		&			1.115					&			2.104			&			0.0683			\\
	1.7		&			0.945					&			1.836			&			0.0148			\\
	1.8		&			0.787					&			1.511			&			0.0126			\\
	2		&			0.572					&			1.081			&			0.00295		\\
	5		&			---						&			---				&			---				\\
	10		&			---						&			---				&			---				\\
	100		&			---						&			---				&			---				\\
	\hline					                
			&					\multicolumn{3}{c}{Long-range Coulomb hole}							\\
												\cline{2-4}
			& Maximum $\Hat{u}^{\rm lr}$		&	Root $\Bar{u}^{\rm lr}$	&	Strength $S^{\rm lr}$	\\
	\hline
	1.6		&			---						&			---				&			---				\\
	1.7		&			2.719					&			3.221			&			0.00334		\\
	1.8		&			2.606					&			3.151			&			0.0267			\\
	2		&			2.737					&			3.382			&			0.0653			\\
	5		&			7.398					&			8.697			&			0.161			\\
	10		&			15.90					&			17.95			&			0.158			\\
	100		&			184.62					&			192.31			&			0.134			\\
\end{tabular}
\end{ruledtabular}
\end{table}

In the weak interaction limit, the only HF solution is the symmetric one (Section \ref{subsec:SHF}) and correlation effects are well-described by the MP approximation (Section \ref{sec:MP}).

Figures \ref{fig:P-MP-R}(a), \ref{fig:P-MP-R}(c) and \ref{fig:P-MP-R}(e) show the Coulomb holes derived from the MP1, MP2 and CI wave functions for three small values of $R = R_1 = R_2$.  For such radii, the MP-based and exact position intracules are very similar, and become identical as $R \to 0$.  The holes are negative for small $u$ and positive for larger $u$, implying that correlation decreases the likelihood of finding the two electrons close together and increases the probability of their being far apart \cite{CoulsonProcPhysSoc1961}.  To illustrate the spatial distributions of the electrons, we have plotted the MP1 Coulomb holes on the surface of a sphere (Fig. \ref{fig:densityplot-MP1}) for the three same values of the $R = R_1 = R_2$.

The evolution of $\Delta P(u)$ with respect to the increase of $R_2$ is shown in Figs. \ref{fig:P-MP-R}(b), \ref{fig:P-MP-R}(d) and \ref{fig:P-MP-R}(f).  As $R_2$ increases, the difference between the MP1 and exact holes decreases and they match perfectly as $R_2 \to \infty$.

Table \ref{tab:DP-MP} shows that the first-order correction reduces the probability of small $u$ values too much, and that the second-order correction partly corrects this, at least for small of values of $R$.  As a consequence, the strength of the MP1 hole is always larger than the true one, but exhibits the right asymptotic behavior for small $R$.

\subsection{Strongly correlated regime}

\begin{figure}
	\begin{center}
	\includegraphics[width=0.48\textwidth]{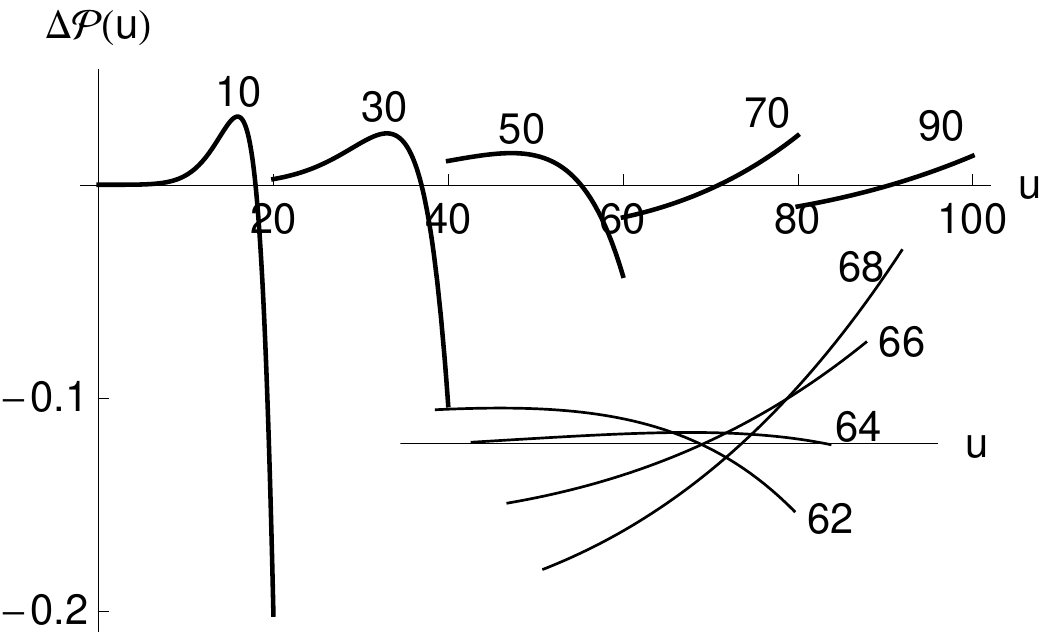}
	\caption{\label{fig:DP-wrt-R2}Holes based on the exact wave function for $R_1 = 10$ and various $R_2$ (from 10 to 90).  The inset graph represents the transition between the AHF and SHF solution ($R_2 = 200/3$).}
	\end{center}
\end{figure}

\begin{figure}
\begin{center}
	\subfigure[~$R_1 = R_2$]{
		\includegraphics[width=0.48\textwidth]{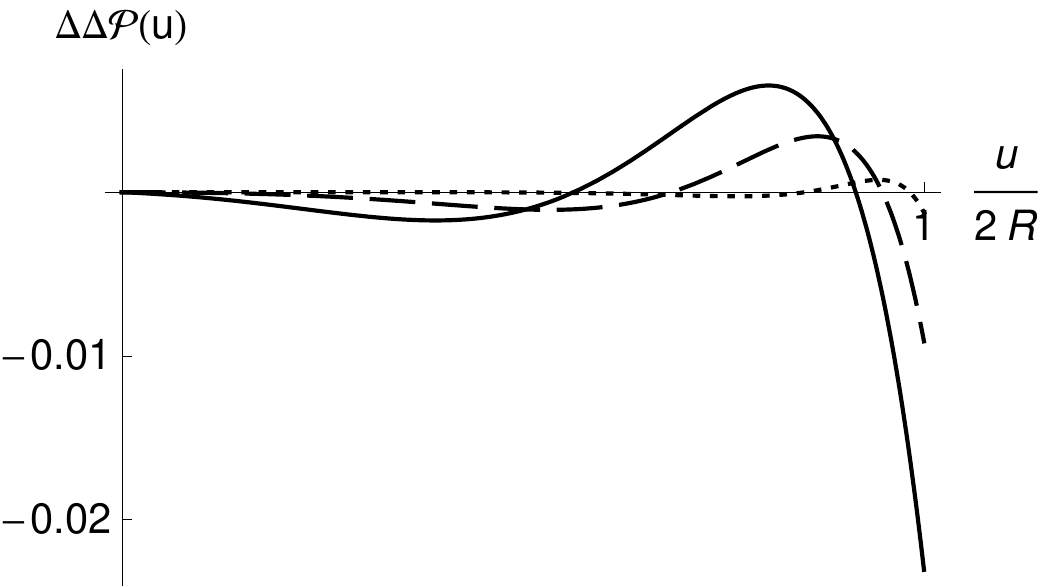}
	}
	\subfigure[~$R_1 = 10$]{
		\includegraphics[width=0.48\textwidth]{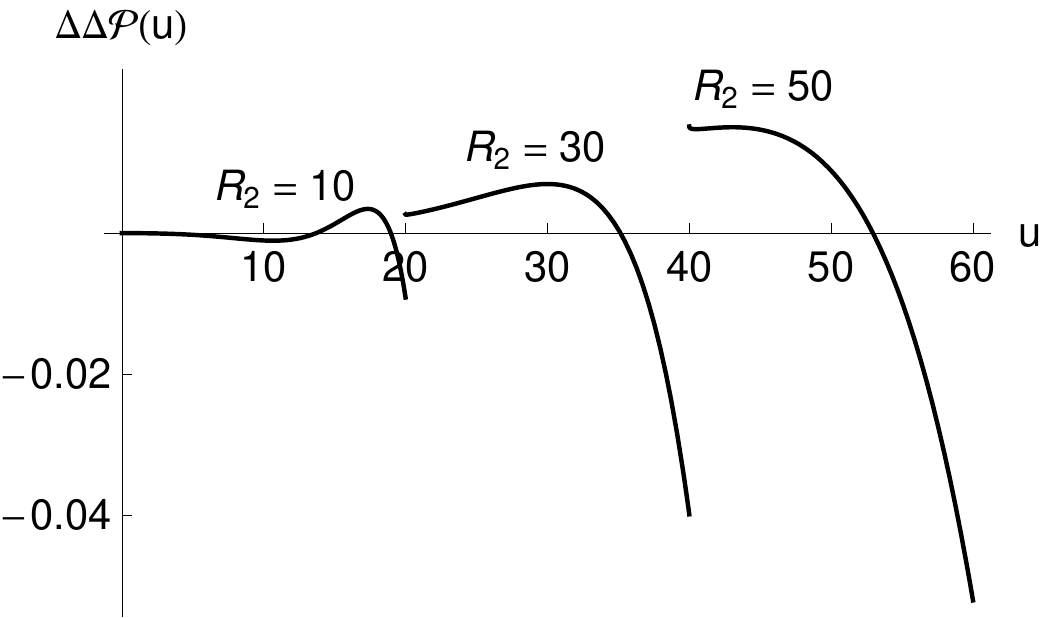}
	}
\caption{\label{fig:DP-LR}
At the top: difference between the exact and the LS holes ($\Delta\Delta\mathcal{P}(u)$) for various $R_1 = R_2$: 5 (solid), 10 (dashed) and 50 (dotted).  At the bottom: $\Delta\Delta\mathcal{P}(u)$ for $R_1 = 10$ and various $R_2$ (10, 30 and 50).
}
\end{center}
\end{figure}

In the strong interaction limit, the Coulomb repulsion dominates the kinetic energy, an AHF solution exists (Section \ref{subsec:AHF}) and the electrons oscillate around their equilibrium positions (Section \eqref{sec:LargeR}).

For $R \gtrsim 1.7$, a secondary Coulomb hole appears in the exact $\Delta \mathcal{P}(u)$ (Fig. \ref{fig:DP-wrt-R}), revealing that correlation \emph{decreases} the probability of finding electrons at large separations.  This implies that the AHF wave function over-localizes the electrons on opposite sides of their spheres, and that correlation then delocalizes them slightly.  Such secondary Coulomb holes are not peculiar to our system; they have also recently been observed in the He atom \cite{PearsonMolPhys2009} and the H$_2$ molecule \cite{PerJCP2009}.

For $R \gtrsim 2$, the primary Coulomb hole disappears completely, leaving only the secondary one (Table \ref{tab:DP-LR} and Fig. \ref{fig:DP-wrt-R}).  The secondary:primary strength ratio is larger than in the He atom and the equilibrium H$_2$ molecule (1-2\%) and resembles that in the H$_2$ molecule at a bond length of 3 a.u. \cite{PerJCP2009}.

Figure \ref{fig:DP-wrt-R2} shows the evolution of the exact hole for $R_1 = 10$ and $R_2$ ranging from 10 to 100.  The secondary hole vanishes when $R_2$ exceeds $R_1^2/R^{\rm crit}$ and the AHF solution collapses to the SHF one.

To compare the holes based on the LS wave function $\Phi^{\omega}$ (Eq. \eqref{Phi-LR}) and the exact one, we have plotted the difference between the exact and the LS holes ($\Delta \Delta \mathcal{P} (u)$ in Fig. \ref{fig:DP-LR}).  For $R = R_1 = R_2$, the agreement between the two holes is fairly good for large $R$ (Fig. \ref{fig:DP-LR}(a)).  For the smaller values of the radius, it shows that the electronic zero-point oscillations tend to over-localize the electrons compared to the exact treatment. However, the secondary Coulomb hole is less pronounced but still present in the LS approximation.  Moreover, one can see that the LS treatment slightly increases the likelihood of finding the two electrons close together.

Figure \ref{fig:DP-LR}(b) reports the modification of $\Delta\Delta\mathcal{P}(u)$ for a fixed value of the first sphere radius ($R_1 = 10$) and various $R_2$ (10, 30 and 50).  When $R_2$ is increasing, the first minimum disappears, and the main effect of the LS approximation is thus to over-localize the electrons on opposite side of the spheres.

\begin{figure}
	\begin{center}
	\includegraphics[width=0.48\textwidth]{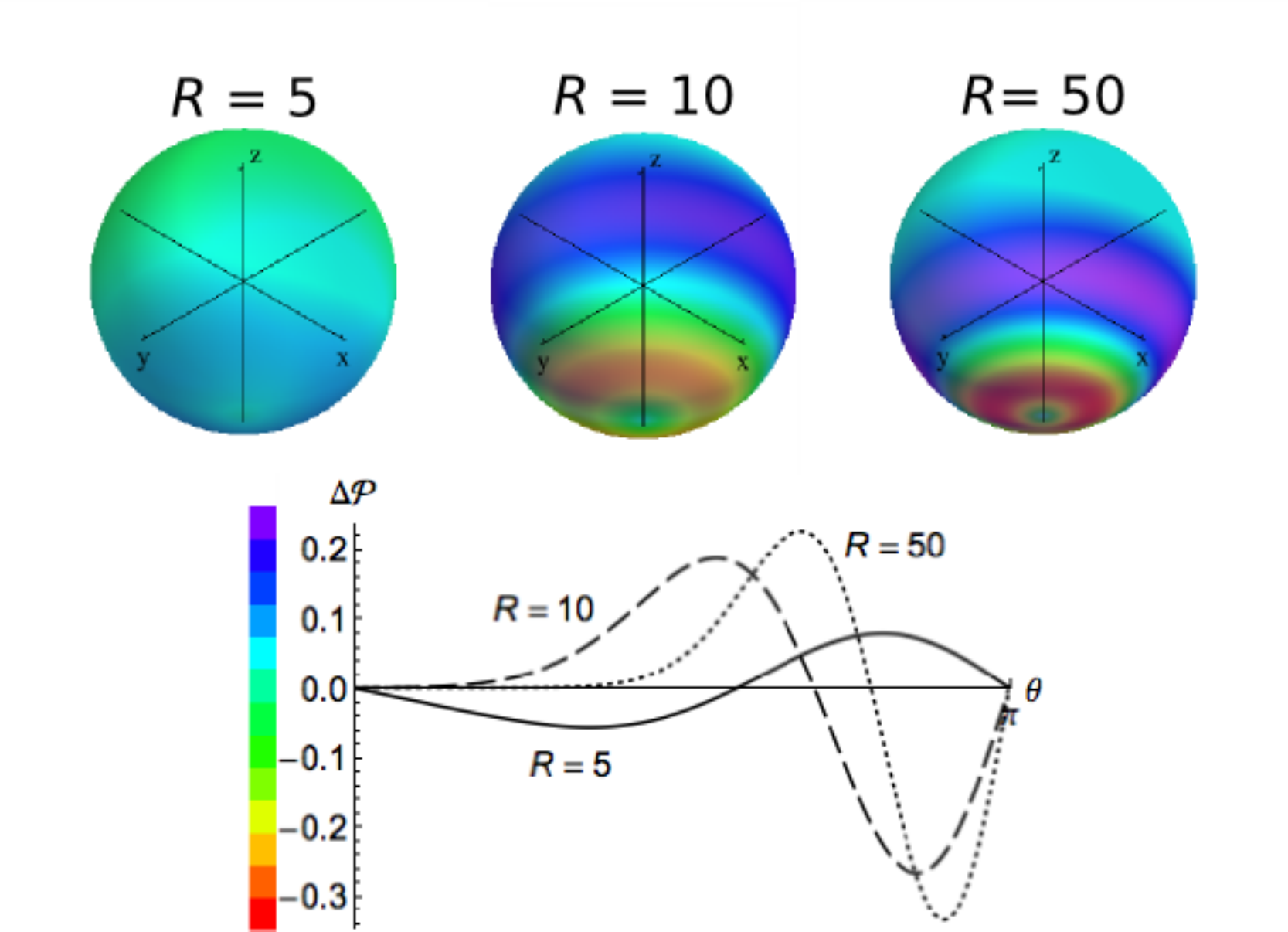}
	\caption{\label{fig:densityplot-LS}LS holes for various $R = R_1 = R_2$ plotted on the surface of a sphere.  The holes with respect to $\theta$ are also represented.}
	\end{center}
\end{figure}

The 2D spatial distribution of the electrons is depicted in Fig. \ref{fig:densityplot-LS}, where we have represented the LS holes for various $R = R_1 = R_2$ (5, 10 and 50) on the surface of a sphere.

\section{Conclusion}
We have performed a comprehensive study of the singlet ground state of two electrons on the surface of spheres of radius $R_1$ and $R_2$.  Symmetric and asymmetric HF solutions show that the symmetry-breaking process occurs only when $R_1 > R^{\rm crit} = 3/2$ and $R_2 < R_1^2/R^{\rm crit}$.  MP2 and MP3 energy corrections reveal that MP theory is appropriate when both radii are small (as previously known) and, also, when $R_2 \gg R_1$.  To derive asymptotic solutions of this problem, we have taken into account, in the harmonic and anharmonic approximations, the zero-point oscillations of the electrons around their equilibrium position.  For any values of $R_2 > R_1$, the near-exact wave function and energy can easily be obtained by a CI expansion based on Legendre polynomials because there is no cusp in the wave function.

A study of the position intracules and Coulomb holes reveals that, as in the helium atom and hydrogen molecule, there is a secondary Coulomb hole in the large-sphere regime.  Indeed, as $R$ increases, the primary hole disappears and only the secondary one remains.  This reflects an overlocalization of the electrons in the asymmetric Hartree-Fock solution.

Our results should be useful for the future development of accurate correlation functionals within density-functional theory \cite{SunJCTC2009, GoriGiorgiJCTC2009, GoriGiorgiPRL2009, SeidlPRA2010} and intracule functional theory \cite{GillPCCP2006, DumontPCCP2007, CrittendenJCP2007a, CrittendenJCP2007b, BernardPCCP2008, PearsonJCP2009} and, also, for understanding secondary Coulomb holes in more complex systems \cite{PearsonMolPhys2009, PerJCP2009}.

\begin{acknowledgments}
P.M.W.G. thanks the NCI National Facility for a generous grant of supercomputer time and the Australian Research Council (Grants DP0771978 and DP0984806) for funding.  We also thank Yves Bernard for fruitful discussions and helpful comments on the manuscript.
\end{acknowledgments}


\begin{thebibliography}{50}
\expandafter\ifx\csname natexlab\endcsname\relax\def\natexlab#1{#1}\fi
\expandafter\ifx\csname bibnamefont\endcsname\relax
  \def\bibnamefont#1{#1}\fi
\expandafter\ifx\csname bibfnamefont\endcsname\relax
  \def\bibfnamefont#1{#1}\fi
\expandafter\ifx\csname citenamefont\endcsname\relax
  \def\citenamefont#1{#1}\fi
\expandafter\ifx\csname url\endcsname\relax
  \def\url#1{\texttt{#1}}\fi
\expandafter\ifx\csname urlprefix\endcsname\relax\def\urlprefix{URL }\fi
\providecommand{\bibinfo}[2]{#2}
\providecommand{\eprint}[2][]{\url{#2}}

\bibitem[{\citenamefont{Loos and Gill}(2009{\natexlab{a}})}]{LoosPRA2009}
\bibinfo{author}{\bibfnamefont{P.-F.} \bibnamefont{Loos}} \bibnamefont{and}
  \bibinfo{author}{\bibfnamefont{P.~M.~W.} \bibnamefont{Gill}},
  \bibinfo{journal}{Phys. Rev. A} \textbf{\bibinfo{volume}{79}},
  \bibinfo{pages}{062517} (\bibinfo{year}{2009}{\natexlab{a}}).

\bibitem[{\citenamefont{Loos and Gill}(2009{\natexlab{b}})}]{LoosPRL2009}
\bibinfo{author}{\bibfnamefont{P.~F.} \bibnamefont{Loos}} \bibnamefont{and}
  \bibinfo{author}{\bibfnamefont{P.~M.~W.} \bibnamefont{Gill}},
  \bibinfo{journal}{Phys. Rev. Lett.} \textbf{\bibinfo{volume}{103}},
  \bibinfo{pages}{123008} (\bibinfo{year}{2009}{\natexlab{b}}).

\bibitem[{\citenamefont{Ezra and Berry}(1982)}]{EzraPRA1982}
\bibinfo{author}{\bibfnamefont{G.~S.} \bibnamefont{Ezra}} \bibnamefont{and}
  \bibinfo{author}{\bibfnamefont{R.~S.} \bibnamefont{Berry}},
  \bibinfo{journal}{Phys. Rev. A} \textbf{\bibinfo{volume}{25}},
  \bibinfo{pages}{1513} (\bibinfo{year}{1982}).

\bibitem[{\citenamefont{Ojha and Berry}(1987)}]{OjhaPRA1987}
\bibinfo{author}{\bibfnamefont{P.~C.} \bibnamefont{Ojha}} \bibnamefont{and}
  \bibinfo{author}{\bibfnamefont{R.~S.} \bibnamefont{Berry}},
  \bibinfo{journal}{Phys. Rev. A} \textbf{\bibinfo{volume}{36}},
  \bibinfo{pages}{1575} (\bibinfo{year}{1987}).

\bibitem[{\citenamefont{Hinde and Berry}(1990)}]{HindePRA1990}
\bibinfo{author}{\bibfnamefont{R.~J.} \bibnamefont{Hinde}} \bibnamefont{and}
  \bibinfo{author}{\bibfnamefont{R.~S.} \bibnamefont{Berry}},
  \bibinfo{journal}{Phys. Rev. A} \textbf{\bibinfo{volume}{42}},
  \bibinfo{pages}{2259} (\bibinfo{year}{1990}).

\bibitem[{\citenamefont{Shytov and Allen}(2006)}]{ShytovPRB2006}
\bibinfo{author}{\bibfnamefont{A.~V.} \bibnamefont{Shytov}} \bibnamefont{and}
  \bibinfo{author}{\bibfnamefont{P.~B.} \bibnamefont{Allen}},
  \bibinfo{journal}{Phys. Rev. B} \textbf{\bibinfo{volume}{74}},
  \bibinfo{pages}{075419} (\bibinfo{year}{2006}).

\bibitem[{\citenamefont{Hohenberg and Kohn}(1964)}]{HohenbergPRB1964}
\bibinfo{author}{\bibfnamefont{P.}~\bibnamefont{Hohenberg}} \bibnamefont{and}
  \bibinfo{author}{\bibfnamefont{W.}~\bibnamefont{Kohn}},
  \bibinfo{journal}{Phys. Rev. B} \textbf{\bibinfo{volume}{136}},
  \bibinfo{pages}{864} (\bibinfo{year}{1964}).

\bibitem[{\citenamefont{Kohn and Sham}(1965)}]{KohnPRA1965}
\bibinfo{author}{\bibfnamefont{W.}~\bibnamefont{Kohn}} \bibnamefont{and}
  \bibinfo{author}{\bibfnamefont{L.}~\bibnamefont{Sham}},
  \bibinfo{journal}{Phys. Rev. A} \textbf{\bibinfo{volume}{140}},
  \bibinfo{pages}{1133} (\bibinfo{year}{1965}).

\bibitem[{\citenamefont{Parr and Yang}(1989)}]{ParrYang}
\bibinfo{author}{\bibfnamefont{R.~G.} \bibnamefont{Parr}} \bibnamefont{and}
  \bibinfo{author}{\bibfnamefont{W.}~\bibnamefont{Yang}},
  \emph{\bibinfo{title}{Density Functional Theory for Atoms and Molecules}}
  (\bibinfo{publisher}{Oxford University Press}, \bibinfo{year}{1989}).

\bibitem[{\citenamefont{Seidl}(2007{\natexlab{a}})}]{SeidlPRA2007b}
\bibinfo{author}{\bibfnamefont{M.}~\bibnamefont{Seidl}},
  \bibinfo{journal}{Phys. Rev. A} \textbf{\bibinfo{volume}{75}},
  \bibinfo{pages}{062506} (\bibinfo{year}{2007}{\natexlab{a}}).

\bibitem[{\citenamefont{Seidl and Gori-Giorgi}(2010)}]{SeidlPRA2010}
\bibinfo{author}{\bibfnamefont{M.}~\bibnamefont{Seidl}} \bibnamefont{and}
  \bibinfo{author}{\bibfnamefont{P.}~\bibnamefont{Gori-Giorgi}},
  \bibinfo{journal}{Phys. Rev. A} \textbf{\bibinfo{volume}{81}},
  \bibinfo{pages}{012508} (\bibinfo{year}{2010}).

\bibitem[{\citenamefont{Seidl et~al.}(2000)\citenamefont{Seidl, Perdew, and
  Kurth}}]{SeidlPRL2000}
\bibinfo{author}{\bibfnamefont{M.}~\bibnamefont{Seidl}},
  \bibinfo{author}{\bibfnamefont{J.~P.} \bibnamefont{Perdew}},
  \bibnamefont{and} \bibinfo{author}{\bibfnamefont{S.}~\bibnamefont{Kurth}},
  \bibinfo{journal}{Phys. Rev. Lett.} \textbf{\bibinfo{volume}{84}},
  \bibinfo{pages}{5070} (\bibinfo{year}{2000}).

\bibitem[{\citenamefont{Moshinsky}(1968)}]{Moshinsky68}
\bibinfo{author}{\bibfnamefont{M.}~\bibnamefont{Moshinsky}},
  \bibinfo{journal}{Am. J. Phys.} \textbf{\bibinfo{volume}{36}},
  \bibinfo{pages}{52} (\bibinfo{year}{1968}).

\bibitem[{\citenamefont{Loos}(2010)}]{LoosPRA2010}
\bibinfo{author}{\bibfnamefont{P.-F.} \bibnamefont{Loos}},
  \bibinfo{journal}{Phys. Rev. A} \textbf{\bibinfo{volume}{81}},
  \bibinfo{pages}{002500} (\bibinfo{year}{2010}).

\bibitem[{\citenamefont{Ezra and Berry}(1983)}]{EzraPRA1983}
\bibinfo{author}{\bibfnamefont{G.~S.} \bibnamefont{Ezra}} \bibnamefont{and}
  \bibinfo{author}{\bibfnamefont{R.~S.} \bibnamefont{Berry}},
  \bibinfo{journal}{Phys. Rev. A} \textbf{\bibinfo{volume}{28}},
  \bibinfo{pages}{1989} (\bibinfo{year}{1983}).

\bibitem[{\citenamefont{Natanson et~al.}(1984)\citenamefont{Natanson, Ezra,
  Delgado-Barrio, and Berry}}]{NatansonJCP1984}
\bibinfo{author}{\bibfnamefont{G.~A.} \bibnamefont{Natanson}},
  \bibinfo{author}{\bibfnamefont{G.~S.} \bibnamefont{Ezra}},
  \bibinfo{author}{\bibfnamefont{G.}~\bibnamefont{Delgado-Barrio}},
  \bibnamefont{and} \bibinfo{author}{\bibfnamefont{R.~S.} \bibnamefont{Berry}},
  \bibinfo{journal}{J. Chem. Phys.} \textbf{\bibinfo{volume}{81}},
  \bibinfo{pages}{3400} (\bibinfo{year}{1984}).

\bibitem[{\citenamefont{Natanson et~al.}(1986)\citenamefont{Natanson, Ezra,
  Delgado-Barrio, and Berry}}]{NatansonJCP1986}
\bibinfo{author}{\bibfnamefont{G.~A.} \bibnamefont{Natanson}},
  \bibinfo{author}{\bibfnamefont{G.~S.} \bibnamefont{Ezra}},
  \bibinfo{author}{\bibfnamefont{G.}~\bibnamefont{Delgado-Barrio}},
  \bibnamefont{and} \bibinfo{author}{\bibfnamefont{R.~S.} \bibnamefont{Berry}},
  \bibinfo{journal}{J. Chem. Phys.} \textbf{\bibinfo{volume}{84}},
  \bibinfo{pages}{2035} (\bibinfo{year}{1986}).

\bibitem[{\citenamefont{Deskevich and Nesbitt}(2005)}]{DeskevichJCP2005}
\bibinfo{author}{\bibfnamefont{M.~P.} \bibnamefont{Deskevich}}
  \bibnamefont{and} \bibinfo{author}{\bibfnamefont{D.~J.}
  \bibnamefont{Nesbitt}}, \bibinfo{journal}{J. Chem. Phys.}
  \textbf{\bibinfo{volume}{123}}, \bibinfo{pages}{084304}
  (\bibinfo{year}{2005}).

\bibitem[{\citenamefont{Deskevich et~al.}(2008)\citenamefont{Deskevich, McCoy,
  Hutson, and Nesbitt}}]{DeskevichJCP2008}
\bibinfo{author}{\bibfnamefont{M.~P.} \bibnamefont{Deskevich}},
  \bibinfo{author}{\bibfnamefont{A.~B.} \bibnamefont{McCoy}},
  \bibinfo{author}{\bibfnamefont{J.~M.} \bibnamefont{Hutson}},
  \bibnamefont{and} \bibinfo{author}{\bibfnamefont{D.~J.}
  \bibnamefont{Nesbitt}}, \bibinfo{journal}{J. Chem. Phys.}
  \textbf{\bibinfo{volume}{128}}, \bibinfo{pages}{094306}
  (\bibinfo{year}{2008}).

\bibitem[{\citenamefont{M{\o}ller and Plesset}(1934)}]{MollerPhysRev1934}
\bibinfo{author}{\bibfnamefont{C.}~\bibnamefont{M{\o}ller}} \bibnamefont{and}
  \bibinfo{author}{\bibfnamefont{M.~S.} \bibnamefont{Plesset}},
  \bibinfo{journal}{Phys. Rev.} \textbf{\bibinfo{volume}{46}},
  \bibinfo{pages}{618} (\bibinfo{year}{1934}).

\bibitem[{\citenamefont{Cizek and Paldus}(1967)}]{CizekJCP1967}
\bibinfo{author}{\bibfnamefont{J.}~\bibnamefont{Cizek}} \bibnamefont{and}
  \bibinfo{author}{\bibfnamefont{L.}~\bibnamefont{Paldus}},
  \bibinfo{journal}{J. Chem. Phys.} \textbf{\bibinfo{volume}{47}},
  \bibinfo{pages}{3976} (\bibinfo{year}{1967}).

\bibitem[{\citenamefont{Paldus and Cizek}(1970)}]{CizekJCP1970}
\bibinfo{author}{\bibfnamefont{L.}~\bibnamefont{Paldus}} \bibnamefont{and}
  \bibinfo{author}{\bibfnamefont{J.}~\bibnamefont{Cizek}}, \bibinfo{journal}{J.
  Chem. Phys.} \textbf{\bibinfo{volume}{52}}, \bibinfo{pages}{2919}
  (\bibinfo{year}{1970}).

\bibitem[{\citenamefont{Seeger and Pople}(1977)}]{SeegerJCP1977}
\bibinfo{author}{\bibfnamefont{R.}~\bibnamefont{Seeger}} \bibnamefont{and}
  \bibinfo{author}{\bibfnamefont{J.}~\bibnamefont{Pople}}, \bibinfo{journal}{J.
  Chem. Phys.} \textbf{\bibinfo{volume}{66}}, \bibinfo{pages}{3045}
  (\bibinfo{year}{1977}).

\bibitem[{\citenamefont{Abramowitz and Stegun}(1972)}]{Abramowitz}
\bibinfo{author}{\bibfnamefont{M.}~\bibnamefont{Abramowitz}} \bibnamefont{and}
  \bibinfo{author}{\bibfnamefont{I.~A.} \bibnamefont{Stegun}},
  \emph{\bibinfo{title}{Handbook of Mathematical Functions with Formulas,
  Graphs and Mathematical Tables}} (\bibinfo{publisher}{Dover publications
  Inc., New-York}, \bibinfo{year}{1972}).

\bibitem[{\citenamefont{Edmonds}(1957)}]{Edmonds}
\bibinfo{author}{\bibfnamefont{A.~R.} \bibnamefont{Edmonds}},
  \emph{\bibinfo{title}{Angular Momentum in Quantum Mechanics}}
  (\bibinfo{publisher}{Princeton University Press}, \bibinfo{year}{1957}).

\bibitem[{\citenamefont{Wigner}(1934)}]{WignerPR1934}
\bibinfo{author}{\bibfnamefont{E.}~\bibnamefont{Wigner}},
  \bibinfo{journal}{Phys. Rev.} \textbf{\bibinfo{volume}{46}},
  \bibinfo{pages}{1002} (\bibinfo{year}{1934}).

\bibitem[{\citenamefont{Alavi}(2000)}]{AlaviJCP2000}
\bibinfo{author}{\bibfnamefont{A.}~\bibnamefont{Alavi}}, \bibinfo{journal}{J.
  Chem. Phys.} \textbf{\bibinfo{volume}{113}}, \bibinfo{pages}{7735}
  (\bibinfo{year}{2000}).

\bibitem[{\citenamefont{Thompson and Alavi}(2002)}]{ThompsonPRB2002}
\bibinfo{author}{\bibfnamefont{D.~C.} \bibnamefont{Thompson}} \bibnamefont{and}
  \bibinfo{author}{\bibfnamefont{A.}~\bibnamefont{Alavi}},
  \bibinfo{journal}{Phys. Rev. B} \textbf{\bibinfo{volume}{66}},
  \bibinfo{pages}{235118} (\bibinfo{year}{2002}).

\bibitem[{\citenamefont{Thompson and Alavi}(2004)}]{ThompsonPRB2004}
\bibinfo{author}{\bibfnamefont{D.~C.} \bibnamefont{Thompson}} \bibnamefont{and}
  \bibinfo{author}{\bibfnamefont{A.}~\bibnamefont{Alavi}},
  \bibinfo{journal}{Phys. Rev. B} \textbf{\bibinfo{volume}{69}},
  \bibinfo{pages}{201302} (\bibinfo{year}{2004}).

\bibitem[{\citenamefont{Thompson and Alavi}(2005)}]{ThompsonJCP2005}
\bibinfo{author}{\bibfnamefont{D.~C.} \bibnamefont{Thompson}} \bibnamefont{and}
  \bibinfo{author}{\bibfnamefont{A.}~\bibnamefont{Alavi}}, \bibinfo{journal}{J.
  Chem. Phys.} \textbf{\bibinfo{volume}{122}}, \bibinfo{pages}{124107}
  (\bibinfo{year}{2005}).

\bibitem[{\citenamefont{Slater}(1960)}]{Slater}
\bibinfo{author}{\bibfnamefont{J.~C.} \bibnamefont{Slater}},
  \emph{\bibinfo{title}{Quantum Theory of Atomic Structures}},
  vol.~\bibinfo{volume}{2} of \emph{\bibinfo{series}{International Series in
  Pure and Applied Physics}} (\bibinfo{publisher}{McGraw-Hill Book Compagny,
  Inc.}, \bibinfo{year}{1960}).

\bibitem[{\citenamefont{Loos and Gill}(2009{\natexlab{c}})}]{LoosJCP2009}
\bibinfo{author}{\bibfnamefont{P.-F.} \bibnamefont{Loos}} \bibnamefont{and}
  \bibinfo{author}{\bibfnamefont{P.~M.~W.} \bibnamefont{Gill}},
  \bibinfo{journal}{J. Chem. Phys.} \textbf{\bibinfo{volume}{131}},
  \bibinfo{pages}{241101} (\bibinfo{year}{2009}{\natexlab{c}}).

\bibitem[{\citenamefont{Dobson et~al.}(2006)\citenamefont{Dobson, White, and
  Rubio}}]{Dobson06}
\bibinfo{author}{\bibfnamefont{J.}~\bibnamefont{Dobson}},
  \bibinfo{author}{\bibfnamefont{A.}~\bibnamefont{White}}, \bibnamefont{and}
  \bibinfo{author}{\bibfnamefont{A.}~\bibnamefont{Rubio}},
  \bibinfo{journal}{Phys. Rev. Lett.} \textbf{\bibinfo{volume}{96}},
  \bibinfo{pages}{073201} (\bibinfo{year}{2006}).

\bibitem[{\citenamefont{Lewin}(1958)}]{Lewin}
\bibinfo{author}{\bibfnamefont{L.}~\bibnamefont{Lewin}},
  \emph{\bibinfo{title}{Dilogarithms and Associated Functions}}
  (\bibinfo{publisher}{London: Macdonald}, \bibinfo{year}{1958}).

\bibitem[{\citenamefont{Seidl et~al.}(1999)\citenamefont{Seidl, Perdew, and
  Levy}}]{SeidlPRA1999a}
\bibinfo{author}{\bibfnamefont{M.}~\bibnamefont{Seidl}},
  \bibinfo{author}{\bibfnamefont{J.~P.} \bibnamefont{Perdew}},
  \bibnamefont{and} \bibinfo{author}{\bibfnamefont{M.}~\bibnamefont{Levy}},
  \bibinfo{journal}{Phys. Rev. A} \textbf{\bibinfo{volume}{59}},
  \bibinfo{pages}{51} (\bibinfo{year}{1999}).

\bibitem[{\citenamefont{Seidl}(1999)}]{SeidlPRA1999b}
\bibinfo{author}{\bibfnamefont{M.}~\bibnamefont{Seidl}},
  \bibinfo{journal}{Phys. Rev. A} \textbf{\bibinfo{volume}{60}},
  \bibinfo{pages}{4387} (\bibinfo{year}{1999}).

\bibitem[{\citenamefont{Seidl}(2007{\natexlab{b}})}]{SeidlPRA2007a}
\bibinfo{author}{\bibfnamefont{M.}~\bibnamefont{Seidl}},
  \bibinfo{journal}{Phys. Rev. A} \textbf{\bibinfo{volume}{75}},
  \bibinfo{pages}{042511} (\bibinfo{year}{2007}{\natexlab{b}}).

\bibitem[{\citenamefont{Gori-Giorgi
  et~al.}(2009{\natexlab{a}})\citenamefont{Gori-Giorgi, Vignale, and
  Seidl}}]{GoriGiorgiJCTC2009}
\bibinfo{author}{\bibfnamefont{P.}~\bibnamefont{Gori-Giorgi}},
  \bibinfo{author}{\bibfnamefont{G.}~\bibnamefont{Vignale}}, \bibnamefont{and}
  \bibinfo{author}{\bibfnamefont{M.}~\bibnamefont{Seidl}}, \bibinfo{journal}{J.
  Chem. Theor. Comput.} \textbf{\bibinfo{volume}{5}}, \bibinfo{pages}{743}
  (\bibinfo{year}{2009}{\natexlab{a}}).

\bibitem[{\citenamefont{Gori-Giorgi
  et~al.}(2009{\natexlab{b}})\citenamefont{Gori-Giorgi, Seidl, and
  Vignale}}]{GoriGiorgiPRL2009}
\bibinfo{author}{\bibfnamefont{P.}~\bibnamefont{Gori-Giorgi}},
  \bibinfo{author}{\bibfnamefont{M.}~\bibnamefont{Seidl}}, \bibnamefont{and}
  \bibinfo{author}{\bibfnamefont{G.}~\bibnamefont{Vignale}},
  \bibinfo{journal}{Phys. Rev. Lett.} \textbf{\bibinfo{volume}{103}},
  \bibinfo{pages}{166402} (\bibinfo{year}{2009}{\natexlab{b}}).

\bibitem[{\citenamefont{Kato}(1957)}]{Kato1957}
\bibinfo{author}{\bibfnamefont{T.}~\bibnamefont{Kato}},
  \bibinfo{journal}{Commun. Pure Appl. Math.} \textbf{\bibinfo{volume}{10}},
  \bibinfo{pages}{151} (\bibinfo{year}{1957}).

\bibitem[{\citenamefont{Coulson and Neilson}(1961)}]{CoulsonProcPhysSoc1961}
\bibinfo{author}{\bibfnamefont{C.~A.} \bibnamefont{Coulson}} \bibnamefont{and}
  \bibinfo{author}{\bibfnamefont{A.~H.} \bibnamefont{Neilson}},
  \bibinfo{journal}{Proc. Phys. Soc. (London)} \textbf{\bibinfo{volume}{78}},
  \bibinfo{pages}{831} (\bibinfo{year}{1961}).

\bibitem[{\citenamefont{Pearson
  et~al.}(2009{\natexlab{a}})\citenamefont{Pearson, Gill, Ugalde, and
  Boyd}}]{PearsonMolPhys2009}
\bibinfo{author}{\bibfnamefont{J.~K.} \bibnamefont{Pearson}},
  \bibinfo{author}{\bibfnamefont{P.~M.~W.} \bibnamefont{Gill}},
  \bibinfo{author}{\bibfnamefont{J.}~\bibnamefont{Ugalde}}, \bibnamefont{and}
  \bibinfo{author}{\bibfnamefont{R.~J.} \bibnamefont{Boyd}},
  \bibinfo{journal}{Mol. Phys.} \textbf{\bibinfo{volume}{07}},
  \bibinfo{pages}{1089} (\bibinfo{year}{2009}{\natexlab{a}}).

\bibitem[{\citenamefont{Per et~al.}(2009)\citenamefont{Per, Russo, and
  Snook}}]{PerJCP2009}
\bibinfo{author}{\bibfnamefont{M.~C.} \bibnamefont{Per}},
  \bibinfo{author}{\bibfnamefont{S.~P.} \bibnamefont{Russo}}, \bibnamefont{and}
  \bibinfo{author}{\bibfnamefont{I.~K.} \bibnamefont{Snook}},
  \bibinfo{journal}{J. Chem. Phys.} \textbf{\bibinfo{volume}{130}},
  \bibinfo{pages}{134103} (\bibinfo{year}{2009}).

\bibitem[{\citenamefont{Sun}(2009)}]{SunJCTC2009}
\bibinfo{author}{\bibfnamefont{J.}~\bibnamefont{Sun}}, \bibinfo{journal}{J.
  Chem. Theor. Comput.} \textbf{\bibinfo{volume}{5}}, \bibinfo{pages}{708}
  (\bibinfo{year}{2009}).

\bibitem[{\citenamefont{Gill et~al.}(2006)\citenamefont{Gill, Crittenden,
  O'Neill, and Besley}}]{GillPCCP2006}
\bibinfo{author}{\bibfnamefont{P.~M.~W.} \bibnamefont{Gill}},
  \bibinfo{author}{\bibfnamefont{D.~L.} \bibnamefont{Crittenden}},
  \bibinfo{author}{\bibfnamefont{D.~P.} \bibnamefont{O'Neill}},
  \bibnamefont{and} \bibinfo{author}{\bibfnamefont{N.~A.}
  \bibnamefont{Besley}}, \bibinfo{journal}{Phys. Chem. Chem. Phys.}
  \textbf{\bibinfo{volume}{8}}, \bibinfo{pages}{15} (\bibinfo{year}{2006}).

\bibitem[{\citenamefont{Dumont et~al.}(2007)\citenamefont{Dumont, Crittenden,
  and Gill}}]{DumontPCCP2007}
\bibinfo{author}{\bibfnamefont{E.~E.} \bibnamefont{Dumont}},
  \bibinfo{author}{\bibfnamefont{D.~L.} \bibnamefont{Crittenden}},
  \bibnamefont{and} \bibinfo{author}{\bibfnamefont{P.~M.~W.}
  \bibnamefont{Gill}}, \bibinfo{journal}{Phys. Chem. Chem. Phys.}
  \textbf{\bibinfo{volume}{9}}, \bibinfo{pages}{5340} (\bibinfo{year}{2007}).

\bibitem[{\citenamefont{Crittenden and Gill}(2007)}]{CrittendenJCP2007a}
\bibinfo{author}{\bibfnamefont{D.~L.} \bibnamefont{Crittenden}}
  \bibnamefont{and} \bibinfo{author}{\bibfnamefont{P.~M.~W.}
  \bibnamefont{Gill}}, \bibinfo{journal}{J. Chem. Phys.}
  \textbf{\bibinfo{volume}{127}}, \bibinfo{pages}{014101}
  (\bibinfo{year}{2007}).

\bibitem[{\citenamefont{Crittenden et~al.}(2007)\citenamefont{Crittenden,
  Dumont, and Gill}}]{CrittendenJCP2007b}
\bibinfo{author}{\bibfnamefont{D.~L.} \bibnamefont{Crittenden}},
  \bibinfo{author}{\bibfnamefont{E.~E.} \bibnamefont{Dumont}},
  \bibnamefont{and} \bibinfo{author}{\bibfnamefont{P.~M.~W.}
  \bibnamefont{Gill}}, \bibinfo{journal}{J. Chem. Phys.}
  \textbf{\bibinfo{volume}{127}}, \bibinfo{pages}{141103}
  (\bibinfo{year}{2007}).

\bibitem[{\citenamefont{Bernard et~al.}(2008)\citenamefont{Bernard, Crittenden,
  and Gill}}]{BernardPCCP2008}
\bibinfo{author}{\bibfnamefont{Y.~A.} \bibnamefont{Bernard}},
  \bibinfo{author}{\bibfnamefont{D.~L.} \bibnamefont{Crittenden}},
  \bibnamefont{and} \bibinfo{author}{\bibfnamefont{P.~M.~W.}
  \bibnamefont{Gill}}, \bibinfo{journal}{Phys. Chem. Chem. Phys.}
  \textbf{\bibinfo{volume}{10}}, \bibinfo{pages}{3447} (\bibinfo{year}{2008}).

\bibitem[{\citenamefont{Pearson
  et~al.}(2009{\natexlab{b}})\citenamefont{Pearson, Crittenden, and
  Gill}}]{PearsonJCP2009}
\bibinfo{author}{\bibfnamefont{J.~K.} \bibnamefont{Pearson}},
  \bibinfo{author}{\bibfnamefont{D.~L.} \bibnamefont{Crittenden}},
  \bibnamefont{and} \bibinfo{author}{\bibfnamefont{P.~M.~W.}
  \bibnamefont{Gill}}, \bibinfo{journal}{J. Chem. Phys.}
  \textbf{\bibinfo{volume}{130}}, \bibinfo{pages}{164110}
  (\bibinfo{year}{2009}{\natexlab{b}}).

\end{thebibliography}
\end{document}